\documentclass[a4paper,11pt]{article}
\usepackage[a4paper,left=0.99in, right=0.99in,top=1.1in, bottom=1.2in]{geometry}

\usepackage{amsmath}
\usepackage{amssymb}
\usepackage{graphicx}
\usepackage{color}
\usepackage{hyperref}
\usepackage{xspace}
\usepackage{color}
\usepackage{ulem}
\usepackage{fixmath}
\usepackage[labelfont=bf,textfont={sl,bf}]{subfig}

\newcommand{\GeV}{\ensuremath{\,\mathrm{GeV}}\xspace}
\newcommand{\TeV}{\ensuremath{\,\mathrm{TeV}}\xspace}
\newcommand{\jet}{\ensuremath{{\,\mathrm{jet}}\xspace}}
\newcommand{\as}{\ensuremath{\alpha_s}}
\newcommand{\aEW}{\ensuremath{\alpha_\text{EW}}}
\newcommand{\nLO}{\nbar \text{LO}\xspace}
\newcommand{\nNLO}{\nbar \text{NLO}\xspace}

\newcommand{\LO}{\text{LO}\xspace}
\newcommand{\NLO}{\text{NLO}\xspace}
\newcommand{\NNLO}{\text{NNLO}\xspace}
\newcommand{\nbar}{\ensuremath{\bar{n}}}
\newcommand{\qbar}{{\bar q}}

\newcommand{\ie}{{\it i.e.}\xspace}

\newcommand{\zzj}{\ensuremath{\ell_1^+ \ell_1^- \, \ell_2^+ \ell_2^- j~}}

\newcommand{\fol}{\ensuremath{\ell_1^+ \ell_1^- \, \ell_2^+ \ell_2^- ~}}
\newcommand{\z}{$Z$ }
\newcommand{\zz}{$ZZ$ }
\newcommand{\zzstar}{$ZZ^*$ }
\renewcommand{\arraystretch}{1.5}
\newcommand{\msbar}{{\overline{\text{MS}}}}
\newcommand{\LS}{{\ensuremath{\,\mathrm{LS}}\xspace}}

\title{\vspace{120pt}
$ZZ$ production at high transverse momenta\\ beyond NLO QCD}
\author{
Francisco~Campanario$^{1,2}$, Michael~Rauch$^2$ and Sebastian~Sapeta$^3$ \\\\
$^1\,${\small\it Theory Division, IFIC, University of Valencia-CSIC,
                E-46980 Paterna, Valencia, Spain}\\
$^2\,${\small\it Institute for Theoretical Physics, Karlsruhe Institute of
                Technology (KIT), Germany} \\
$^3\,${\small\it CERN PH-TH, CH-1211, Geneva 23, Switzerland}
}

\date{}

\begin{document}
\maketitle 
\flushbottom

\thispagestyle{empty}

\vspace{-36em}
\begin{flushright}
CERN-PH-TH-2015-079\;\;\\
FTUV-15-0329\;\;\\
IFIC-15-22\;\;\\
KA-TP-07-2015\;\;\\
LPN15-023\;\;
\end{flushright}
\vspace{26em}

\begin{abstract}
We study the production of the four-lepton final state $\ell^+ \ell^-
\ell^+ \ell^-$, predominantly produced by a pair of electroweak $Z$
bosons, $ZZ$.
Using the LoopSim method, we merge NLO QCD results for $ZZ$ and $ZZ$+jet and
obtain approximate NNLO predictions for $ZZ$ production. The exact gluon-fusion
loop-squared contribution to the $ZZ$ process 
is also included. 
On top of that, we add to our merged sample the gluon-fusion $ZZ$+jet contributions from the
gluon-gluon channel, which is formally of N$^3$LO and provides approximate results at NLO for the gluon-fusion mechanism.
The predictions are obtained with the
VBFNLO package and include the leptonic decays of the \z bosons
with all off-shell and spin-correlation effects, as well as virtual
photon contributions.
We compare our predictions with existing results for the total inclusive cross
section at NNLO and find a very good agreement.  Then, we present results for
differential distributions for two experimental setups, one used in searches for
anomalous triple gauge boson couplings, the other in Higgs analyses in the
four charged-lepton final state channel.
We find that the approximate NNLO corrections are large, reaching up to 20\% at
high transverse momentum of the \z boson or the leading lepton, and are not
covered by the NLO scale uncertainties. 
Distributions of the four-lepton invariant mass are, however, 
stable with respect to QCD corrections at this order.

%
\end{abstract}

\newpage
\tableofcontents

\section{Introduction}
\label{sec:intro}
The production of a pair of electroweak vector bosons constitutes an
excellent avenue to test the electroweak sector of the Standard Model
(SM) at the LHC.  This class of processes provides for example
information on the non-abelian structure of the Lagrangian. Of
particular relevance is the production of a pair of $Z$ bosons, where the
underlying gauge structure of $SU(2)_L \otimes U(1)_Y$ predicts that
tri-linear couplings are actually absent at tree-level in the SM.
Furthermore, there is a contribution from an $s$-channel Higgs boson
resonance, produced in gluon-fusion (GF) via a heavy-quark mediated
effective Higgs-gluon coupling. This part allows to perform interference
studies and off-shell Higgs width
measurements~\cite{Kauer:2012hd,Caola:2013yja,Campbell:2013una,Gainer:2014hha,Englert:2014aca,Cacciapaglia:2014rla}.

By the experiments, this process is measured indirectly via $Z$ bosons
decaying to a pair of charged leptons or neutrinos each. In this
article, we will focus on the four charged-lepton final state
\begin{equation} 
pp \rightarrow \ell_1^+ \ell_1^- \, \ell_2^+ \ell_2^- +
X.
\end{equation}
In the past years, the ATLAS and CMS collaborations have provided a rich
collection of measurements with increasing accuracy of the four lepton
signal, and limits on the trilinear anomalous $ZZ\gamma$ and $ZZZ$ vertices have
been deduced~\cite{Aad:2012awa,Khachatryan:2014dia}.

The next-to-leading order (NLO) QCD corrections to $ZZ$ production were
first computed in Refs.~\cite{Mele:1990bq,Ohnemus:1990za} for on-shell
production, and including the leptonic decays and spin correlations
in Refs.~\cite{Campbell:1999ah,Dixon:1999di}. They turn out to be sizable, of
around 50$\%$, and exhibit a relevant phase-space dependence. In
differential distributions, much larger corrections up to the order of
10 can appear. This makes the approximation of correcting the LO
differential distribution by the total K-factor (the ratio of the NLO
over the LO total cross section), an unreliable estimate of the true NLO
differential cross section and can severely underestimate its size.  The
origin of the large magnitude of the corrections is twofold. At NLO, new
partonic sub-processes appear, including those with enhanced gluonic
parton distribution functions (PDFs). This explains in part the size of
the total K-factor.  On the other hand, some new topologies appear for
the first time only at NLO. Among them is a topology with a soft or collinear
boson emission from a quark or anti-quark, which results in an $\alpha_s
\alpha_{EW} \ln^2 (p_{T,j} /m_V)$ enhancement for a number of
observables~\cite{Rubin:2010xp,Campanario:2012fk,Campanario:2013wta}.

The one-loop gluon-induced corrections $gg \to \ell_1^+ \ell_1^- \,
\ell_2^+ \ell_2^- + X $ are currently known only at LO. They were first
reported for on-shell production in
Refs.~\cite{Dicus:1987dj,Glover:1988fe}. Results including the leptonic
decays ~\cite{Matsuura:1991pj,Zecher:1994kb} are also available and
studies in the framework of Higgs measurements have also been carried
out
(e.g. Refs.~\cite{Kauer:2012hd,Englert:2014aca,Adam:2008aa,Binoth:2008pr}). 
Formally, they contribute to $ZZ$ production only at
next-to-next-to-leading order (NNLO) QCD, but due to the large gluon
PDFs at the LHC, their numerical impact is larger than this naive
counting of coupling constants suggests. Depending on the selected cuts,
their contribution ranges from a few percent up to ten percent. This
prediction suffers from large scale uncertainties. However, results at
NLO QCD for the gluonic contributions are expected to be available soon
-- the real corrections, $ZZj$ production, and the virtual two-loop
corrections are already known~\cite{Campanario:2012bh,Caola:2015ila}.

The NLO electroweak corrections for on-shell $ZZ$ production have been
computed in Refs.~\cite{Bierweiler:2013dja,Baglio:2013tva}. They yield
only a modest contribution, ranging from the few percent level for
integrated cross sections up to 10-20$\%$ for high-$p_T$ observables.

At NNLO QCD, further new partonic channels and new topologies
contribute. Thus, to match with the expected experimental precision, it
is mandatory to assess the size of these NNLO corrections, not only at
the total cross section level, but also for the differential
distributions. $ZZj$ production at NLO QCD provides the mixed
real-virtual and the double real ${\cal O} (\alpha_s^2)$ contributions
to the NNLO results. They were first computed in
Ref.~\cite{Binoth:2009wk} for on-shell production and account for the
new sub-processes and the new topologies appearing for the first time at
NNLO. Thus, they are expected to provide the dominant contribution to
the total NNLO prediction for selected observables.

The NLO QCD corrections to the double real-emission process, $pp \to
\ell_1^+ \ell_1^- \, \ell_2^+ \ell_2^- \, jj + X $ have been reported
recently~\cite{Campanario:2014ioa} with corrections around 10$\%$. The
size of the corrections is relatively mild, if adequate central scales
are chosen, due to the absence of new channels and phase-space regions
opening up at this order, although the uncertainty from varying the
factorization and renormalization scale gets greatly reduced.

The two-loop virtual corrections for off-shell $ZZ$ production have been
presented in several
publications~\cite{Gehrmann:2014bfa,Caola:2014iua,Gehrmann:2015ora,
vonManteuffel:2015msa} and the NNLO QCD result for the inclusive total cross
section has been reported recently~\cite{Cascioli:2014yka}. The size of the NNLO
corrections compared to the NLO result is about 10\%. Up to date, no fully
differential NNLO predictions are available in the literature.

In this article, we employ the LoopSim method~\cite{Rubin:2010xp,
LoopSim} together with the NLO predictions for $ZZ$ and $ZZj$ production
and the LO ones for GF-$ZZ$ and GF-$ZZj$, calculated by the Monte Carlo
program VBFNLO~\cite{Arnold:2008rz,Arnold:2011wj,Baglio:2014uba}. From this procedure, we obtain merged samples
that, in certain regions of phase space, are expected to account for the
dominant part of the NNLO QCD corrections to the $ZZ$ production process.
The abovementioned procedure has been used recently for other diboson production
processes~\cite{Campanario:2012fk, Campanario:2013wta}, as well as V+jet
processes~\cite{Maitre:2013wha}, resulting in corrections ranging from
30-100$\%$ for selected distributions.  Such sizable corrections should
be taken into account in experimental analyses.

The article is organized as follows: In Section~\ref{sec:theo}, the
details of the theoretical framework of our calculation are given. In
section~\ref{sec:num}, first, we compare our predictions with the
existing ones presented in Ref.~\cite{Cascioli:2014yka} for on-shell $Z$
pair production at the integrated cross section level at
NNLO. Afterwards, differential distributions are also presented for two
set of cuts -- in Section~\ref{sec:incl}, the setup of the ATLAS and CMS experimental
analyses on $ZZ$ production has been closely followed, while in
Section~\ref{sec:hig}, Higgs search cuts are imposed, following the CMS
analysis of Ref.~\cite{Chatrchyan:2012ufa}.  Finally, in
Section~\ref{sec:su}, we present our summary and conclusions.

\section{Theoretical Framework}
\label{sec:theo}

\begin{figure}
\includegraphics[width=\textwidth]{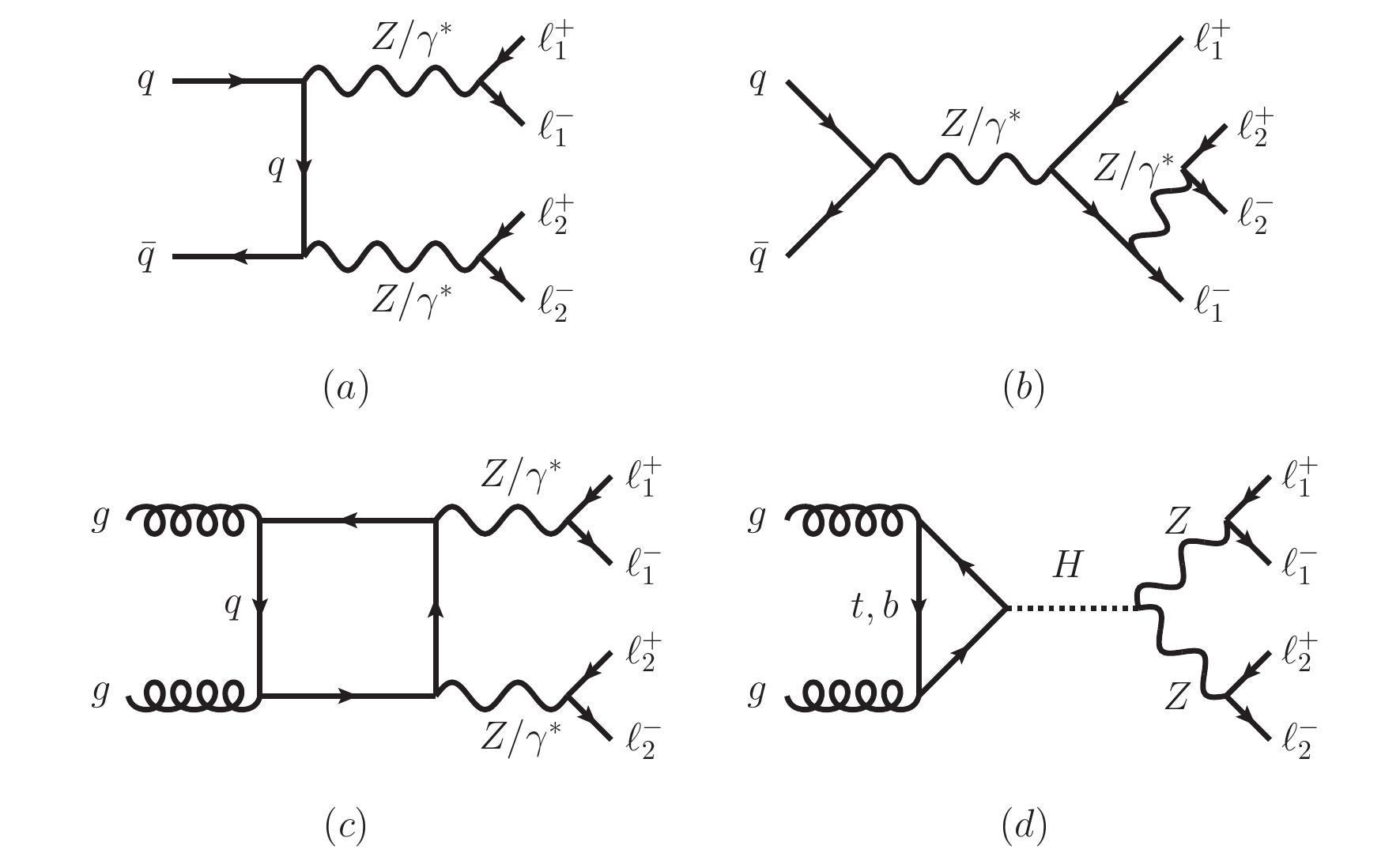}
\caption{Representative Feynman diagrams for the different production
processes contributing to $pp \to \ell_1^+ \ell_1^- \, \ell_2^+
\ell_2^-$. \textit{Top:} $q\bar{q}$ processes appearing at tree-level.
\textit{Bottom:} GF-initiated processes appearing at the
one-loop-squared level.}
\label{fig:feynmanZZ}
\end{figure}

Production of the four-lepton final state happens at leading order
mainly via the quark--anti-quark $t/u$-channel diagram (see
Fig.~\ref{fig:feynmanZZ}~$(a)$ for a representative Feynman diagram). The bulk of the
contribution comes from on-shell $ZZ$ production, as in that case the
electroweak coupling in the $Z$ decay gets effectively replaced by the
corresponding branching ratio, which is about $10\%$ for a $Z$ boson
decaying into a pair of any charged leptons. Instead of $Z$ bosons, also
diagrams appear where these are replaced by virtual photons, denoted as
$\gamma^*$ in the following. These typically yield lepton pairs with
invariant masses much smaller than the $Z$ mass. Their overall
contribution strongly depends on the lepton cuts imposed on the final
state. Typical experimental invariant mass windows for $Z$ bosons
have a lower bound of 66 GeV, which reduces the $\gamma^*$
contribution to a negligible level. 

Another possibility to produce this four-lepton final state is
production of a single vector boson $ V \in (Z, \gamma^*)$, which in
turn undergoes a four-body decay into the final-state leptons. An
example Feynman diagram is depicted again in Fig.~\ref{fig:feynmanZZ}
$(b)$. 
This $s$-channel contribution is also sub-dominant in the SM, since there
are no tree-level tri-linear gauge couplings and selection cuts on
the final-state leptons suppress these contributions due to the
limited phase space to simultaneously produce the two intermediate 
vector bosons close to their mass shell.


Finally, there are one-loop-squared GF diagrams that can also generate
the same four-lepton final-state. These can either proceed via an $s$-channel
Higgs resonance, which subsequently decays into \fol 
and is produced via an effective Higgs-gluon-gluon coupling mediated by loops of heavy-quarks,
predominantly the top quark. The dominant contribution to the $H \to \fol$
decay comes from an intermediate $ZZ^*$ system, with one on-shell
and one off-shell intermediate vector boson.
The other possibility is a continuum
production of two, potentially off-shell, $Z/\gamma^*$ bosons through a
quark-loop box diagram. For typical inclusive cuts,
the bulk of the GF contributions originates from on-shell $ZZ$
production, similar to the t/u-channel diagrams. 
Production via an $s$-channel vector boson
resonance is forbidden by gauge invariance and the Landau-Yang
theorem~\cite{Landau:1948kw,Yang:1950rg}.

For simplicity, if not stated otherwise, we will refer to the whole
process by $ZZ$ production, although we will consider all off-shell
effects, non-resonant diagrams and spin correlations of the four-lepton
final state. Also, adding the contributions from virtual photons
$\gamma^*$ is implicitly understood if not mentioned otherwise. 

\subsection[$ZZ$ and $ZZj$ production in VBFNLO]
           {$\mathbold{ZZ}$ and $\mathbold{ZZj}$ production in VBFNLO}

Our calculation relies on the following ingredients.
The NLO QCD corrections for the $ZZ$ and the $ZZj$ production processes, as well as
the LO one-loop gluon-fusion-induced $gg\to $ \fol and $gg \to $ \zzj
contributions are obtained from the VBFNLO package. The NLO QCD
corrections to $ZZj$ were included first for this study,
and, in the following, we give some details of the methodology used to
compute them to make this work self-contained. We follow closely the
strategies used for $\ell \nu_\ell \gamma \gamma j$ production~\cite{Campanario:2011ud}.

We use the spinor-helicity amplitude method and the effective current
approach~\cite{Hagiwara:1988pp,Campanario:2011cs} to factorize the
electroweak part of the system, containing the leptonic tensor, from the
QCD amplitude. 
We first generate the generic $q\bar{q} \to V_1 V_2 j $
amplitudes, where, here and in the following, $V$
denotes either a, possibly virtual, $Z^{(*)}$ boson or a virtual
photon $\gamma^*$. Additionally, we also calculate the contribution $q
\bar{q} \to V j $. 
Then the leptonic decays $ V_i \to \ell_i^+ \ell_i^-$ and
$V \to \fol$ are attached via effective currents. Note that, in this way,
all the off-shell effects and spin correlations are taken into
account. All possible flavor and cross-related sub-processes are
computed from these generic amplitudes. 

At NLO, we need to compute the virtual and the
real corrections. To regularize the ultraviolet (UV) and infrared (IR)
divergences, we use dimensional regularization~\cite{'tHooft:1972fi} and
the anti-commuting prescription of
$\gamma_5$~\cite{Chanowitz:1979zu}. We employ the Catani-Seymour
algorithm~\cite{Catani:1996vz} to explicitly cancel the IR
divergences prior to the phase-space integration. 

To evaluate the scalar integrals, we follow the prescription of
Refs.~\cite{'tHooft:1978xw,Bern:1993kr,Dittmaier:2003bc} and for the 
one-loop tensor coefficients, we employ the Passarino-Veltman reduction
formalism~\cite{Passarino:1978jh} up to the box level, but avoiding the
explicit appearance of Gram
determinants~\cite{Campanario:2011cs,Hahn:2006qw}, and the formalism of
Refs.~\cite{Binoth:2005ff,Denner:2005nn}, with the notation laid out in
Ref.~\cite{Campanario:2011cs} for pentagon integrals.

Several checks have been applied to the virtual amplitudes, computed
with the package described in Ref.~\cite{Campanario:2011cs}, among them, the factorization of
the poles, gauge invariance, and parametrization invariance~\cite{Campanario:2011cs}. 
For the virtual contributions with closed fermion loops, 
we set the mass of all the fermions to zero and check that the poles add to zero. Additionally, we have cross-checked both the
continuum and Higgs resonance graphs with a second
implementation based on FeynArts and
FormCalc~\cite{Hahn:2000jm,Hahn:2006zy,Rauch:2008fy,Gross:2014kda}. 
On the amplitude level, there are 10-14 digits of agreement for each
case. The integrated leading
order and real emission contributions have been checked against
Sherpa~\cite{Gleisberg:2008ta} and agreement at the per mille level has been found.
Additionally, we have cross-checked that our predictions agree at the 
per mille level with the ones provided in MCFM~\cite{Campbell:2010ff}.

A similar strategy is used for the calculation of $ZZ$ at NLO
QCD and the LO gluon fusion $gg \to $ \fol and $gg \to $ \zzj
contributions. We include the leptonic decays via effective currents
using the spin-helicity formalism and all off-shell effects including
Higgs graphs and photons are taken into account. More details on the
treatment of the challenging numerical instabilities appearing in $gg
\to $ \zzj can be found in Ref.~\cite{Campanario:2012bh}.

For the final-state leptons, we consider $Z$ decays into the first two
generations, \ie electrons and muons. We neglect contributions from
Pauli interference due to identical particles in the final-state, which
is below the per mille level. Therefore, all four possible decay
combinations ($V_1$, $V_2$ decaying into $e$ or $\mu$ each) give the
same result and need to be produced only once at the generation level.
As we will see later, the cuts are different for same-flavor and
different-flavor states, and in the Higgs setup case also for electrons
and muons, hence after imposing analysis cuts, the contributions of the
four combinations will differ. For the $Z$ resonance, we employ a
modified version of the complex-mass scheme~\cite{Denner:1999gp} where the
weak mixing angle is kept real. 
We work in the five-flavor scheme and use the $\msbar$
renormalization of the strong coupling constant, with the top quark
decoupled from the running of $\alpha_s$. 
Diagrams with a final state top quark pair, which would appear in the
real emission part of $ZZj$, are considered a separate, experimentally
distinguishable process and are therefore discarded. Virtual top-loop
contributions in contrast are included in our calculation. We
consider a massless bottom-quark, $m_b = 0$, except for
the closed bottom-quark loop amplitudes, where it is set
to its pole mass value. The latter is important to correctly account for
the negative interference between top- and bottom-quark loops in the
effective Higgs-gluon coupling, while for the continuum diagrams the
difference between choosing a massless or massive bottom quark is
numerically small.

\subsection{Computing the dominant part of NNLO with LoopSim}

For the calculation of approximate NNLO results for $pp\to$ \fol 
production, we use the LoopSim method~\cite{Rubin:2010xp, LoopSim}. 
This approach allows us to merge, in a consistent manner,
the $ZZ$@NLO and $ZZj$@NLO samples, provided by
VBFNLO, yielding a result which is simultaneously NLO accurate 
both for the $ZZ$ and $ZZj$ production processes. This merged
sample is expected to provide the dominant part of the NNLO QCD
correction to $ZZ$ production in phase space regions where additional jet
radiation becomes important. 

The NLO $ZZj$ predictions provide the double-real and the mixed real-virtual
contributions at NNLO of the $ZZ$ result. 
They are divergent upon integration over the phase space of the real partons
and need to be supplemented with double-virtual contributions to yield the
finite NNLO result for $ZZ$ production.
LoopSim constructs approximate versions of such double-virtual terms by
utilizing the fact that their structure of divergences has to match exactly
that of the higher multiplicity contributions.
This procedure guarantees finiteness of the combined result,
while providing a dominant part of the NNLO result for a number of relevant
observables.
In the process of determining the exact divergent terms of the two-loop
corrections, some finite pieces are generated, which are not guaranteed to match
the exact constant part of the two-loop diagram.
They are, however, proportional to the LO born
kinematics and are therefore negligible for distributions which receive sizable
corrections at NLO.

To link the two programs, we have made use of an interface~\cite{Campanario:2012fk}
which consists, on one hand, of an extension of the VBFNLO program that writes
down events in the Les Houches event (LHE)~\cite{Alwall:2006yp} format at NLO and, on
the other hand, of a class in the LoopSim library that reads and
processes the LHE events.

The LoopSim method proceeds in the following steps. 
First, an underlying structure for each  NLO $ZZj$ event is determined with
the help of the Cambridge/Aachen (C/A)~\cite{Dokshitzer:1997in,Wobisch:1998wt}
algorithm, as implemented in FastJet~\cite{Cacciari:2005hq, FastJet}, with a
certain radius $R_{\text{LS}}$. This step establishes the sequence of emissions
in the input event.
For simplicity, we combine each pair of oppositely-charged leptons to a
virtual vector boson $V$ and LoopSim processes diagrams at this level.
Thereby, we use information from the event generation step to
identify the leptons connected by a continuous fermion line.
For $s$-channel-type events, which are dominated by contributions where
only a single boson is attached to the QCD part of the amplitude, one
might consider to also reflect this in the LoopSim part and combine the
four leptons into a single particle. The question would then be how to
define the transition between the two regions. As this contribution
is strongly suppressed by the cuts applied later, we do not
pursue this further, but instead always combine the four leptons into
two $Z$ bosons.
In the next step, the underlying hard structure of the event is determined by
working through the $ij\to k$ recombinations in order of decreasing hardness,
defined by the $k_t$ algorithm measure~\cite{Kt, Ellis:1993tq}.
The first $n_b$ particles associated with the hardest merging are marked as
``Born''. The number of Born particles is fixed by the number of outgoing
particles in the LO event and is equal to two for the case of $ZZ$ production.
At NNLO, the Born particles can be either both vector bosons, a boson and a
parton, or two partons.
The remaining particles, which are not marked as ``Born'', are then ``looped''
by finding all possible ways of recombining them with the emitters. This step
generates approximate one- and two-loop diagrams. 

In the next step, a double counting between the approximate one-loop events
generated by LoopSim and the exact one-loop events coming from the NLO sample
with lower multiplicity is removed. 
This is done by generating the one-loop diagrams from the tree level events
first, and then using them to generate all possible one and two-loop events.
This set is then subtracted from the full result, which amounts to
removing the two-loop diagrams with both loops simulated by LoopSim and leaving
only those with one exact and one simulated loop.
To satisfy unitarity, these diagrams are assigned a weight equal to the original
event times a prefactor $(-1)^\text{number of loops}$. This guarantees that the
sum of the weights of the LoopSim-generated events is zero~\cite{Rubin:2010xp}.
For a fully inclusive observable, the integrated cross section of
our approximated NNLO result would be equal to the NLO one (hence the pure $\aEW^4\as^2$
contributions would vanish). However, in realistic situations, with finite fiducial volumes
or in the case of differential distributions, some of the events generated by
LoopSim are removed by cuts or reshuffled within histograms, which results in
non-vanishing $\aEW^4\as^2$
correction and that leads to a genuine, approximate
NNLO correction.

The jet C/A and $k_t$ algorithms mentioned above depend on the
radius $R_\LS$, which is a parameter of the LoopSim method. 
The smaller the value of $R_\LS$, the more likely the particles are recombined
with the beam, the larger $R_\LS$, the more likely they are recombined
together. The value of $R_\LS$ is irrelevant for collinear and soft radiation.
It affects only the wide angle (or hard) emissions where the mergings between
particles $i$ and $j$ compete with mergings with the beam.
In our study, we shall use $R_\LS = 1$, and we shall vary it by $\pm 0.5$. 
The $R_\LS$ uncertainty will therefore account for the part of the LoopSim
method which is related to attributing the emission sequence and the underlying
hard structure of the events.

In order to distinguish our predictions with simulated loops from those with
exact loop diagrams, we denote the approximate loops by $\nbar$, as opposed to N
used for the exact ones. 
With that notation, for processes whose contributions start at tree level like
$q\qbar \to ZZ$, 
$\nLO$ denotes the correction with simulated one-loop diagrams, and $\nNLO$ is a
result with exact one-loop and simulated two-loop contributions. 
However, for processes that start contributing only at one-loop, like $gg\to
ZZ$, $\nLO$ denotes the correction with respect to that first, non-trivial
result. Hence, it is formally an N$^3$LO contribution with respect to
the full process of $ZZ$ production.

The GF contribution formally first contributes at NNLO, and consequently
we also include it in our merged $\nNLO$ sample generated with LoopSim. 
Due to the large gluonic PDFs, this process can contribute relevantly
despite the $\alpha_s^2$ suppression. Hence, and since it is gauge
invariant on its own, by now a common approach in the literature is to
add this contribution already to the NLO results. We follow this
convention, but make the addition explicit by using the label
``NLO+LO-GF'' in this case.
Additionally, as mentioned before, we also merge the real radiation
process GF-$ZZj$ computed at LO to the GF-$ZZ$ result, yielding a
contribution appearing only at N$^3$LO. Our results will in general also
include this contribution, where we label the full results as
``\nNLO+\nLO-GF''.
In summary, the GF contribution is always implicitely understood to be
included at the corresponding order given by the power counting of coupling
constants, in particular LO-GF in the \nNLO result. If additional GF
contributions are added, this is made explicit in the label.

\section{Numerical Results}
\label{sec:num}

\subsection{Comparison with inclusive NNLO calculation}
\label{sec:int}
\begin{table}
\begin{center}
\renewcommand{\arraystretch}{1.4}
\begin{tabular*}{0.95\textwidth}{@{\extracolsep{\fill}}|l@{\hspace{0.8cm}}|l@{}l|@{\extracolsep{\fill}}}
\hline
$\sigma_{\LO}$ [pb] & 
  5.0673(4) ${}^{+1.6\%}_{-2.7\%}$ &(\textit{Ref.~\cite{Cascioli:2014yka}}: 5.060 ${}^{+1.6\%}_{-2.7\%}$)\\ \hline   
$\sigma_{\NLO}$ [pb] & 
  7.3788(10) ${}^{+2.8\%}_{-2.3\%}$ &(\textit{Ref.~\cite{Cascioli:2014yka}}: 7.369 ${}^{+2.8\%}_{-2.3\%}$)\\ \hline   
$\sigma_{\NLO+\text{LO-GF}}$ [pb] & 
  7.946(3) ${}^{+4.2\%}_{-3.2\%}$ & \\ \hline
$\sigma_{\NNLO}$ [pb] & 
  & (\textit{Ref.~\cite{Cascioli:2014yka}}: 8.284 ${}^{+3.0\%}_{-2.3\%}$)\\ \hline   
$\sigma_{\nNLO}$ [pb] & 
  8.103(5) ${}^{+4.7\%}_{-2.6\%}$ $(\mu)$ \quad ${}^{+0.8\%}_{-0.6\%}$ ($R_{LS}$) & \\ \hline   
$\sigma_{\nNLO+\nLO\text{-GF}}$ [pb] & 
  8.118(5) ${}^{+4.7\%}_{-2.6\%}$ $(\mu)$ \quad ${}^{+0.8\%}_{-0.6\%}$ ($R_{LS}$) & \\ \hline   
\end{tabular*}
\end{center}
\caption{Comparison with Ref.~\cite{Cascioli:2014yka} of total cross sections 
for on-shell $ZZ$ production at the LHC running at $\sqrt{s}=8\, \text{TeV}$.
The errors in brackets are the statistical error
from Monte Carlo integration, while the percentages give the scale
variation error, obtained from changing $\mu_F$ and $\mu_R$
independently within the range $[\frac12 m_Z; 2 m_Z]$, where the ratio
$\mu_F/\mu_R$ is constrained to stay within $[\frac12;2]$. 
For the $\nNLO$ results we additionally
give the error due to a variation of $R_{\LS}$ between $0.5$ and $1.5$.
}
\label{tab:csNNLOcomp}
\end{table}
We start by comparing the results obtained with LoopSim+VBFNLO with the
calculation of the inclusive NNLO cross section of
Ref.~\cite{Cascioli:2014yka}. Here, in contrast to the rest of the
paper, we use the settings as those of Ref.~\cite{Cascioli:2014yka}, \ie the $Z$ bosons are on-shell
and do not decay, hence no cut is placed on the final state. Also, all
numerical values of masses and couplings are taken from there, namely
$m_W=80.399\GeV$, $m_Z=91.1876\GeV$,
$G_F=1.16639\times10^{-5}\GeV^{-2}$, $m_t=173.2\GeV$, $m_H=125\GeV$ and
$\mu_R=\mu_F=m_Z$. As PDFs, the MSTW 2008 set~\cite{Martin:2009iq} is chosen,
evaluated at each corresponding order.
Since our matrix elements include all off-shell effects and spin
correlations in our prediction of \fol production, for this comparison,
we have to modify our code. 
All the $s$-channel contributions, like the example diagram in
Fig.~\ref{fig:feynmanZZ}~$(b)$, or the $t/u$-channel contributions with $\gamma^*$
in the intermediate state are set to zero. Hence, the only contribution comes
from $t/u$-channel diagrams of Fig.~{\ref{fig:feynmanZZ}} with on-shell $Z$
bosons. For the GF part, both Higgs and continuum diagrams contribute,
but for the latter we also have to remove all diagrams with virtual photons.
In the phase-space
generator, the (Breit-Wigner) distributions for the invariant mass of
the $Z$ bosons are replaced by $\delta$-distributions at the $Z$ pole
mass. As leptonic decays of the $Z$ bosons are still simulated
internally, we finally need to account for this by dividing the result
by $\text{BR}(Z\to \ell_i^+\ell_i^-)^2$.
The resulting cross sections at 8~TeV are shown in
Table~\ref{tab:csNNLOcomp}.

At LO and NLO, agreement at the per mille level is found. The NLO
integrated K-factor, defined as the ratio of NLO/LO predictions, is 1.46.
For comparison, we also show the result of adding the GF contribution,
formally NNLO, to the NLO result, evaluating both with NNLO PDFs. This
is the best currently available estimate without requiring the
evaluation of two-loop diagrams or merging different jet multiplicities.
We see that these give an additional 7.7\% contribution to the NLO cross
section, or a total K factor of 1.57 comparing NLO+GF to the LO result.
Additionally, they increase the scale variation uncertainty
significantly.
The latter comes from the fact that the GF contribution is the lowest order
accuracy result for the $gg$~channel.  

With respect to NLO, the total NNLO
correction, computed in Ref.~\cite{Cascioli:2014yka} and quoted in the fourth
row of Table~\ref{tab:csNNLOcomp}, is about $12\%$. Hence, the GF contribution
provides the leading part of those, namely 60\%.
Our approximate \nNLO and \nNLO+\nLO-GF results are shown in the last two rows
of Table~\ref{tab:csNNLOcomp}. Their overall agreement with the full NNLO result
is good with the difference at the level of 2$\%$ only. 
This is not a priori guaranteed by our method and is consistent with the 
assumption of a LO $\times$ $\alpha_s^2$ effect coming from the
genuine finite pieces of the exact two-loop virtual amplitudes. 
These terms are not properly determined by our method, but they are
covered by the remaining scale uncertainty.
The scale uncertainty of our \nNLO result is similar to NLO+GF, and not
reduced like for the NNLO result, since the LoopSim method does not
attempt to reconstruct higher-order terms proportional to the scale
dependence in order not to underestimate the variation, although,
technically, this would be possible. 
%

\subsection{Differential distributions}
\label{sec:diff}

In the following sections, results at the LHC at $\sqrt{s} =8$ \TeV will be given
for two different sets of cuts. In Section~\ref{sec:incl}, we closely
follow the ATLAS and CMS experimental analyses on $ZZ$ production, while
in Section~\ref{sec:hig}, we impose Higgs search cuts, following the CMS
analysis of Ref.~\cite{Chatrchyan:2012ufa}.  Below, we
describe the common settings. 

As input parameters, we use
\begin{align}
m_Z &= 91.1876 \GeV\,, & G_F &= 1.16637 \times 10^{-5}\GeV^{-2}\,, \nonumber\\
m_W &= 80.398  \GeV\,, & \alpha_\text{em}^{-1} &= 132.3407     \,, \nonumber\\
m_H &= 125     \GeV\,, & \sin^2(\theta_W) &= 0.22265\,, \\ 
\Gamma_Z &= 2.508\GeV\,, &  \Gamma_H &= 0.004017 \GeV  \,. \nonumber 
\end{align}
The mass of the top and bottom quarks, which run in the closed fermion loops,
are set to
\begin{align}
m_t &= 172.4 \GeV\,, & m_b = 4.855 \GeV \ .
\end{align}
All other quarks, including external bottom quarks, are taken as
massless. 

The jets are defined with the anti-$k_t$ 
algorithm~\cite{Cacciari:2008gp}, as implemented in
FastJet~\cite{Cacciari:2005hq, Cacciari:2011ma}, with the radius $R=0.4$. 
Independently of the order of a prediction, we use the NNLO
MSTW2008~\cite{Martin:2009iq} PDF set, provided by the
LHAPDF~\cite{LHAPDF} implementation with $\alpha_s(m_Z)= 0.11707$.

At fixed order in perturbation theory, the cross section depends on the
renormalization and factorization scale. As central values for both of those
scales, we choose
the scalar sum of the transverse energy of the system
\begin{equation}
\mu_{F,R}= \mu_0 = \frac{1}{2} \left( 
\sum  p_{T,\text{partons}} + 
\sqrt{p_{T,V_1}^2+m_{V_1}^2}+
\sqrt{p_{T,V_2}^2+m_{V_2}^2} 
\right) \,,
\label{eq:ren}
\end{equation}
where $p_{T,V_{1,2}}$ and $m_{V_{1,2}}$ are the transverse momenta and invariant
masses of the recombined, opposite-signed charged lepton pairs, respectively.
The scale uncertainty is obtained by varying simultaneously the factorization
and renormalization scale by a factor two around the central scale.
Additionally, to assess the uncertainties associated with the recombination
method used by LoopSim, we show the uncertainty bands associated with variations
of $\pm 0.5$ around $R_{\text{LS}}=1$.

\subsubsection[$ZZ$ analysis]{$\mathbold{ZZ}$ analysis}
\label{sec:incl}

In the analysis of $ZZ$ production, we use the following cuts, inspired largely
by the ATLAS paper~\cite{Aad:2012awa}. The settings of the corresponding
CMS analysis~\cite{Khachatryan:2014dia} are comparable.
The transverse momenta and pseudorapidities of leptons, as well as those of jets
(for observables exclusive with respect to jet activity), and the distance
between leptons and leptons and jets are required to stay in the following
fiducial volume:
\begin{align}
p_{t,\ell} &> 20 \GeV\,,      &  |\eta_{\ell}| &<2.5\,,        \nonumber \\
\label{eq:cutsincl}
p_{t,\jet} &> 25 \GeV\,,      &  |\eta_{\jet}| &<4.5\,,                  \\
\Delta R_{\ell,\jet}& >0.3\,, & \Delta R_{\ell,\ell}  &>0.2\,. \nonumber
\end{align}

To reconstruct the $Z$ bosons from the leptons, we employ the following
algorithm. First, all invariant-mass pairs of same-flavor and
opposite-sign lepton pairs are formed. If all leptons are of the same
generation, there are in total four possibilities, while for different
generations only two exist. 
The invariant-mass pair closest to the physical $Z$ boson mass is labeled $Z_1$
and it is required to satisfy the cut $66 \GeV < m_{\text{inv},Z_1} < 116 \GeV$,
otherwise the event is discarded.
If the second pair of leptons, which we denote as $Z_2$, falls into the same
mass window, the event is labelled as $ZZ$, if not, it is called a
$ZZ^\star$ event, provided that $m_{Z_2} > 20 \GeV$. If the latter is
not satisfied, the event is rejected. Hence, our two selection types can be summarized as
\begin{equation}
  \begin{array}{lc}
    ZZ   \text{ selection: }   & m_{Z_1}, m_{Z_2} \in (66, 116) \GeV\,,  \\
    ZZ^* \text{ selection: } & m_{Z_1} \in (66, 116) \GeV\,,\
    m_{Z_2} \in (20, 66)  \cup (166,m_{Z,\max}) \GeV\,,
  \end{array}
  \label{eq:mZselections}
\end{equation}
where $m_{Z,\max}$ is the maximal mass that can be obtained for a given energy
of the system of the incoming partons.
Note that the terminology adopted for our study differs slightly from that of
Ref.~\cite{Aad:2012awa}, where $ZZ^\star$ was used for the union of both
categories defined in Eq.~(\ref{eq:mZselections}).

\begin{table}
\begin{center}
\renewcommand{\arraystretch}{1.4}
\begin{tabular*}{\textwidth}{@{\extracolsep{\fill}}|l@{\hspace{0.5cm}}|ll|@{\extracolsep{\fill}}}
\hline
& \multicolumn{1}{c}{$ZZ$} & \multicolumn{1}{c|}{$ZZ^*$} \\
\hline
$\sigma_{\LO}$ [fb] & 
  9.394(9) ${}^{+2.2\%}_{-3.1\%}$ & 
  1.0134(16) ${}^{+1.2\%}_{-1.9\%}$ \\ \hline 
$\sigma_{\NLO}$ [fb] & 
  12.057(19) ${}^{+1.6\%}_{-1.0\%}$ &
  1.314(3) ${}^{+2.0\%}_{-1.5\%}$ \\ \hline 
$\sigma_{\NLO+\text{LO-GF}}$ [fb] & 
  12.929(19) ${}^{+3.4\%}_{-2.4\%}$ &
  1.365(3) ${}^{+3.0\%}_{-2.2\%}$ \\ \hline
$\sigma_{\nNLO}$ [fb] & 
  13.15(8) ${}^{+3.3\%}_{-2.3\%}$ $(\mu)$ ${}^{+0.8\%}_{-0.6\%}$ ($R_{LS}$) &
  1.417(12) ${}^{+2.0\%}_{-1.4\%}$ $(\mu)$ ${}^{+0.8\%}_{-0.7\%}$ ($R_{LS}$) \\ \hline 
$\sigma_{\nNLO+\nLO\text{-GF}}$ [fb] & 
  13.15(8) ${}^{+3.3\%}_{-2.3\%}$ $(\mu)$ ${}^{+0.9\%}_{-0.7\%}$ ($R_{LS}$) &
  1.427(12) ${}^{+2.3\%}_{-1.6\%}$ $(\mu)$ ${}^{+0.9\%}_{-0.7\%}$ ($R_{LS}$) \\ \hline 
\end{tabular*}
\end{center}
\caption{Inclusive cross sections at $\sqrt{s}=8\, \text{TeV}$ for the process $pp\to\fol$ using the
cuts of Eq.~(\ref{eq:cutsincl}), separated into $ZZ$ and $ZZ^*$ event
categories defined in Eq.~(\ref{eq:mZselections}). The errors in brackets are the statistical error
from Monte Carlo integration, while the percentages give the scale
variation error, obtained from varying $\mu = \mu_F = \mu_R
\in [\frac12 \mu_0; 2 \mu_0]$, with $\mu_0$ given by Eq.~\ref{eq:ren}. 
For the $\nNLO$ results, we additionally
give the error due to a variation of $R_{LS}$ between $0.5$ and $1.5$.
}
\label{tab:csincl}
\end{table}
In Table~\ref{tab:csincl}, we present the inclusive cross section
at different levels of accuracy. As one can see, the overall behaviour
is similar to what we have already observed for total on-shell $ZZ$
production in the previous subsection. The GF contribution gives a
significant correction to the NLO result of about $+7.2\%$ for the $ZZ$
case, while for $ZZ^*$ it is only $+3.9\%$. In both cases the scale
dependence is strongly increased. The additional integrated \nNLO
corrections are modest with $1.7\%$ and $3.8\%$ for the $ZZ$ and $ZZ^*$ cases,
respectively. The dependence on
the LoopSim-Parameter $R_{\LS}$ is clearly smaller than the remaining
scale variation error.

A more important aspect for our method are, however, differential
distributions, where the effects can be much larger. To this, we will
turn next.


\begin{figure}[t]
  \centering
  \includegraphics[width=0.45\columnwidth]{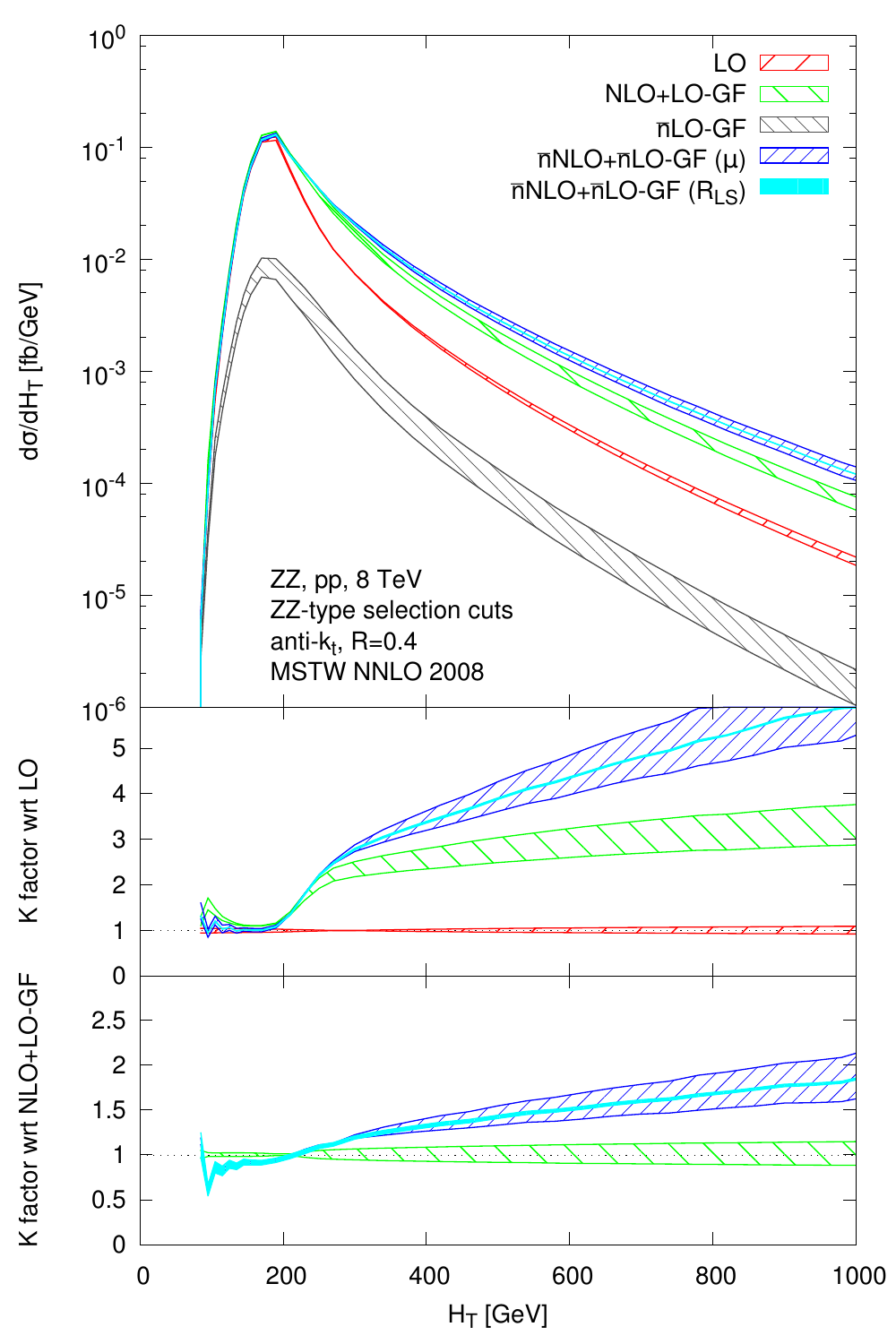}
  \hfill
  \includegraphics[width=0.45\columnwidth]{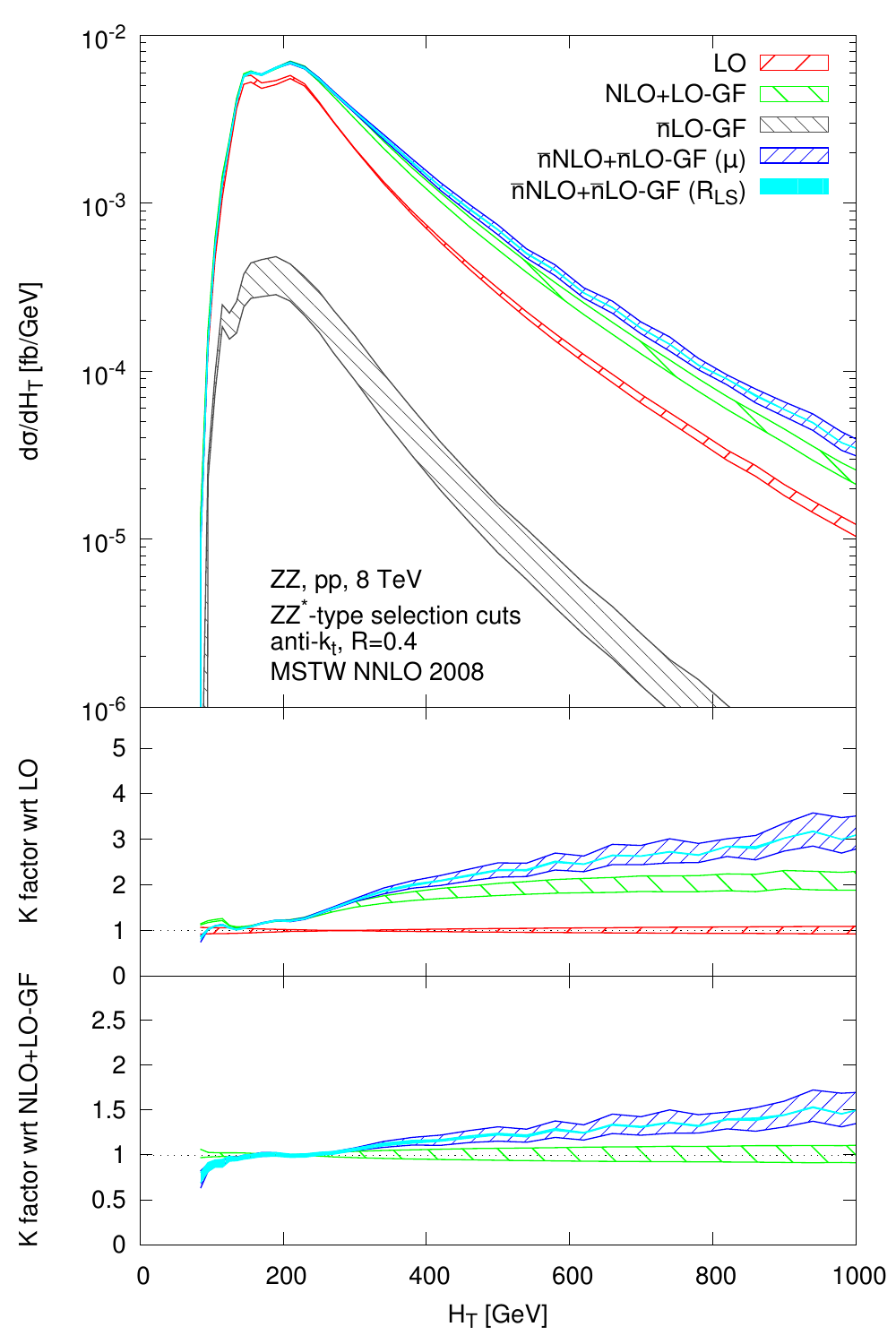}
  \caption{  
  Differential cross sections and K factors for the effective mass observable $H_T$,
  defined in Eq.~(\ref{eq:HT}), for the LHC at $\sqrt{s}=8\, \text{TeV}$. 
  The left and the right plot correspond to the $ZZ$ and $ZZ^*$ selections
  defined in Eq.~(\ref{eq:mZselections}).
  The bands correspond to varying $\mu_F=\mu_R$ by factors 1/2 and 2 around the
  central value from Eq.~(\ref{eq:ren}). 
  The cyan solid bands give the
  uncertainty related to the $R_\text{LS}$ parameter varied between 0.5 and 1.5.
  The distribution is a sum of contributions from same-flavor decay
  channels ($4e$ and $4\mu$) and the different-flavor channel ($2e2\mu$).
  }
  \label{fig:HT-standard}
\end{figure}

\begin{figure}[t!]
  \centering
  \includegraphics[width=0.80\columnwidth]{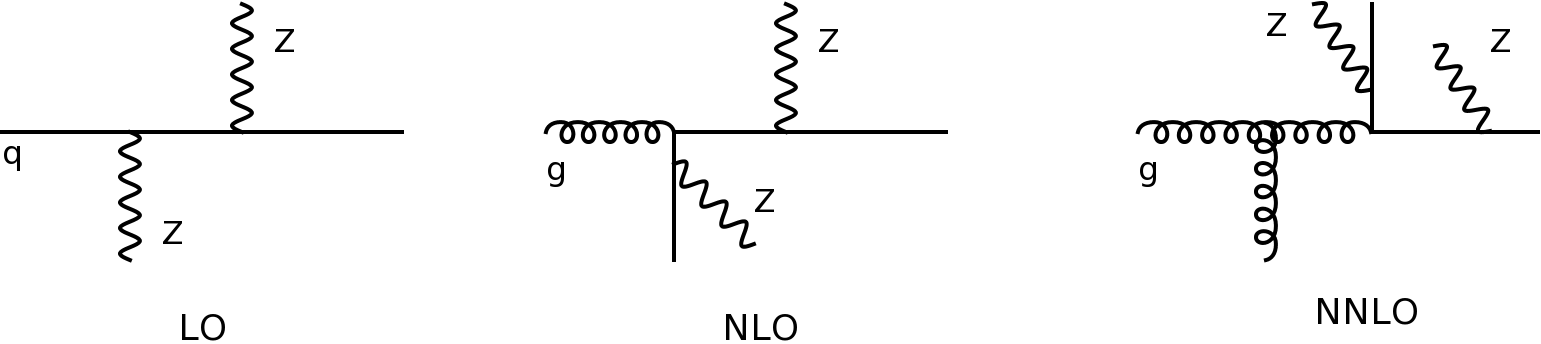}
  \caption{Example diagrams contributing to $ZZ$ production at LO, NLO and NNLO.}
  \label{fig:ZZdiagrams}
\end{figure}

\begin{figure}[t]
  \centering
 \includegraphics[width=0.45\columnwidth]{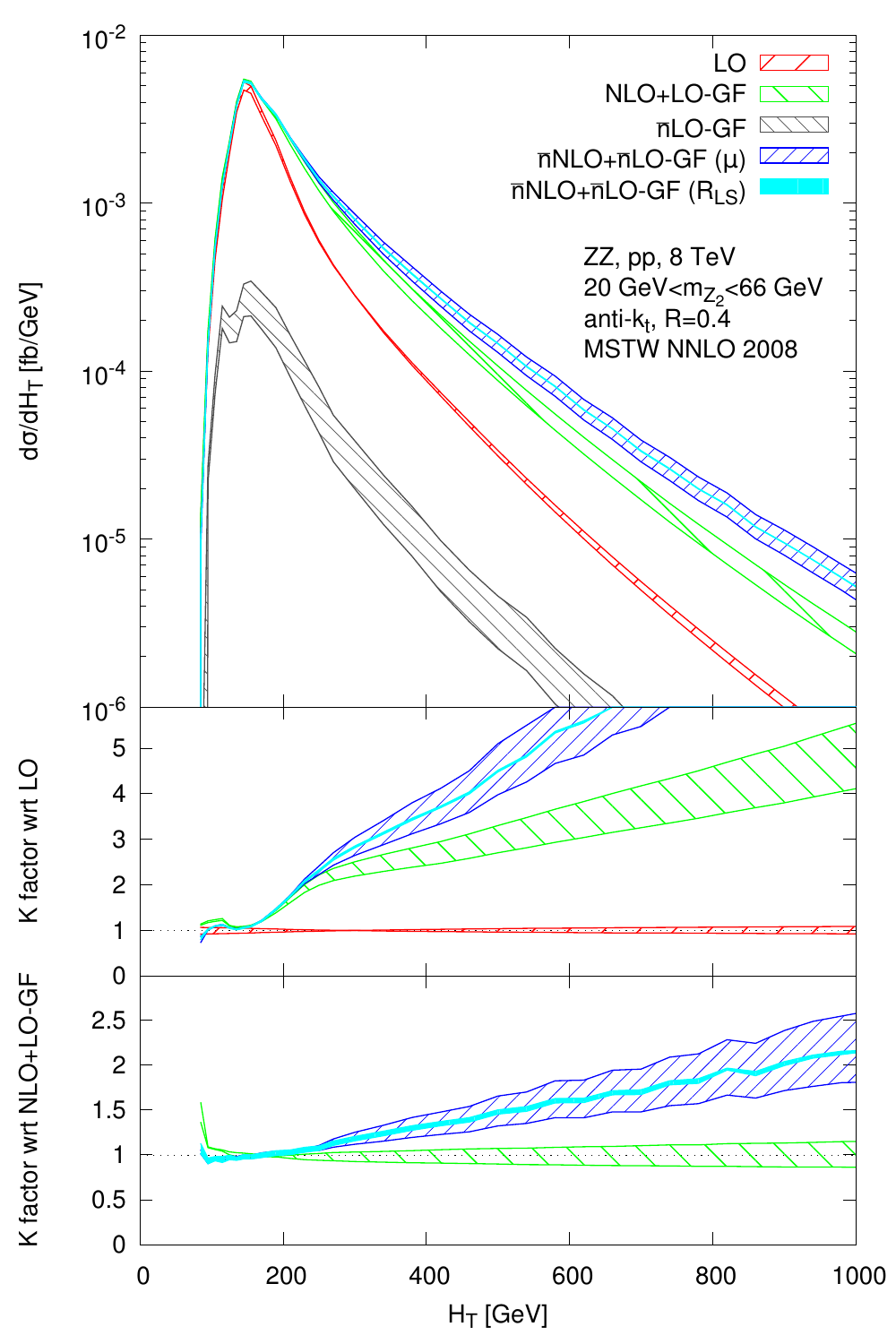}
  \hfill
  \includegraphics[width=0.45\columnwidth]{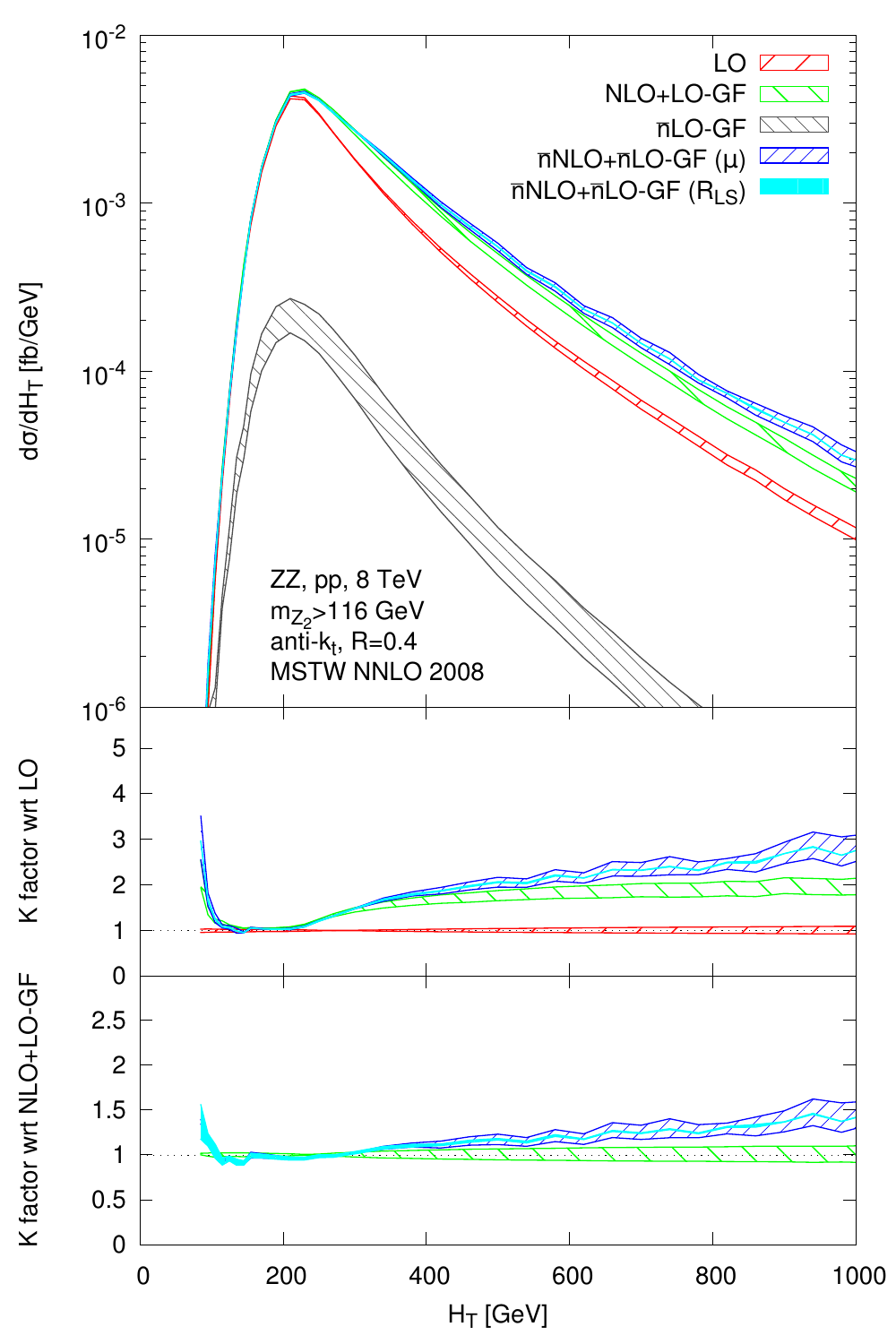}
  \caption{  
  Differential cross sections and K factors for the effective mass observable $H_T$,
  defined in Eq.~(\ref{eq:HT}) for the LHC at $\sqrt{s}=8\, \text{TeV}$. 
  The left plot corresponds to the part of \zzstar selection,
  Eq.~(\ref{eq:mZselections}), where $20 < m_Z < 66 \GeV$, and the right plot to the 
  part of \zzstar selection with $m_Z > 116 \GeV$.
  Other details are as in Fig.~\ref{fig:HT-standard}.
  }
  \label{fig:HT-mass-regions}
\end{figure}

Fig.~\ref{fig:HT-standard} shows distributions of the effective mass observable,
$H_T$, defined as a scalar sum of the transverse momenta of leptons and jets, and
missing transverse energy
\begin{equation}
H_{T} = \sum  p_{T,\text{jets}} + \sum  p_{T,l} +
E_{T,\text{miss}}\,.
\label{eq:HT}
\end{equation}
The left panel corresponds to the $ZZ$ and the right panel to the \zzstar types
of cuts. 
In the former case, the K factor is very large, both at NLO and
at \nNLO. In the case of \zzstar selection, the K factor is visibly
smaller. The leading correction to this observable at NLO comes from
configurations, shown in the middle diagram of Fig.~\ref{fig:ZZdiagrams}, with
one of the bosons emitted collinearly and the other with large transverse
momentum recoiling against a hard jet $p_{T,\jet} \simeq
p_{T,Z}$~\cite{Frixione:1992pj}, which results in a dependence given by
\begin{equation}
  \frac{d\sigma}{d\Omega} \propto
  \ln \frac{p_{T,\jet}^2}{m_Z^2}\,.
  \label{eq:log-enh}
\end{equation}
A similar enhancement occurs at NNLO, where both $Z$ bosons are allowed to be soft or
collinear and the result is dominated by the dijet type configurations shown in
Fig.~\ref{fig:ZZdiagrams} (right).
This explains both why the K factors grow with transverse momentum and shows
that the rate of this growth depends on the selection of the vector boson mass.
 
In Fig.~\ref{fig:HT-mass-regions}, we split the result of
Fig.~\ref{fig:HT-standard}~(right) into the two separate mass regions for the off-shell
bosons: $20 < m_{Z_2} < 66 \GeV$~Fig.~\ref{fig:HT-mass-regions}~(left) and $m_{Z_2} >
116 \GeV$~Fig.~\ref{fig:HT-mass-regions}~(right).  
We see that the NLO and the \nNLO K factors are much larger in the former case,
as putting a smaller mass in Eq.~(\ref{eq:log-enh}) leads to a stronger logarithmic
enhancement.  We also see that, at large $H_T$, it is the right plot of
Fig.~\ref{fig:HT-mass-regions} that contributes more to the sum shown in
Fig.~\ref{fig:HT-standard}~(right), in terms of absolute values.  This comes
from the fact that the born cross section is already significantly larger in
this case.
Therefore, Fig.~\ref{fig:HT-mass-regions}~(right), with larger $m_Z$ values,
which based on Eq.~(\ref{eq:log-enh}) results in a lower K factor,
dominates the total result of Fig.~\ref{fig:HT-standard}~(right). That is why
the enhancement is smaller there, as compared to
Fig.~\ref{fig:HT-standard}~(left), where only the mass region around the $Z$ peak is included.

\begin{figure}[t]
  \centering
  \includegraphics[width=0.45\columnwidth]{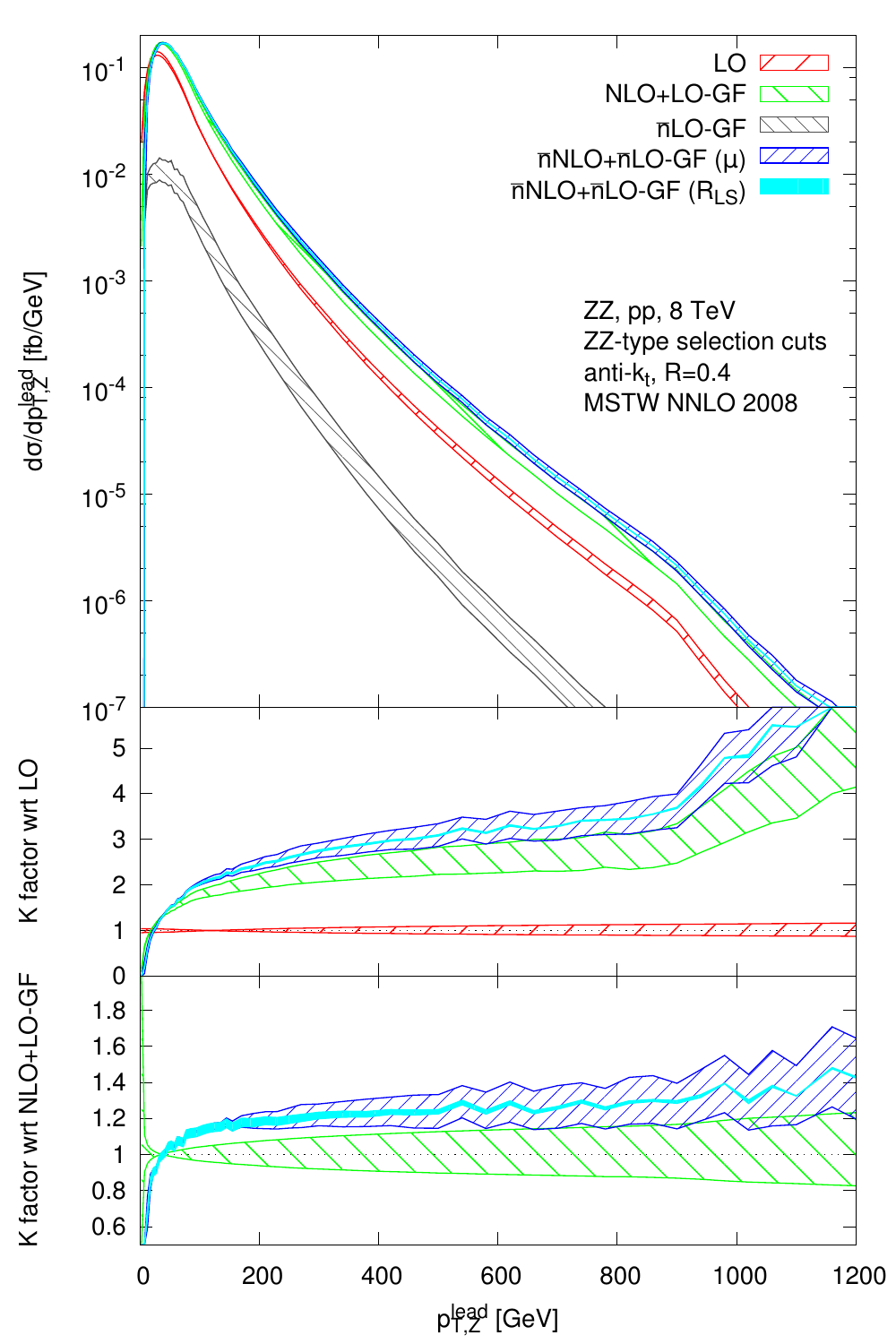}
  \hfill
  \includegraphics[width=0.45\columnwidth]{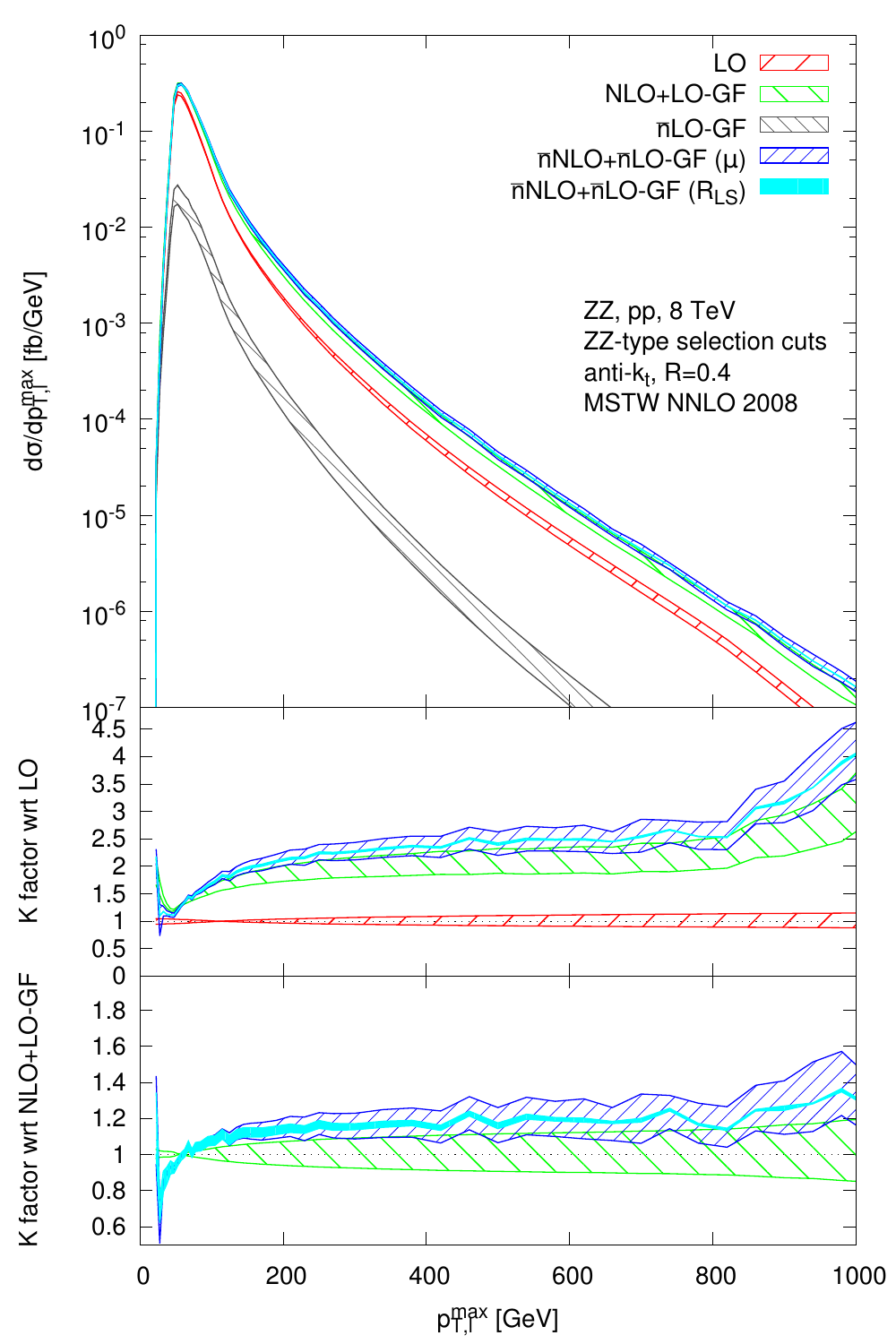}
  \caption{  
  Differential cross sections and K factors for the transverse momentum of the
  leading $Z$ boson~(left) and the leading lepton~(right) for the \zz type
  selection of Eq.~(\ref{eq:mZselections}) at the LHC with $\sqrt{s}=8\,
  \text{TeV}$. 
  All details as in Fig.~\ref{fig:HT-standard}.
  }
  \label{fig:pTlmax-ptZlead}
\end{figure}

Let us now turn to Fig.~\ref{fig:pTlmax-ptZlead}, where we show distributions of
the transverse momentum of the leading $Z$ boson and that of the leading lepton
for the $ZZ$ selection. We see that, in both cases, the \nNLO correction is 
significant, reaching up to 20\%  with respect to the
NLO result at high $p_T$.
This magnitude of \nNLO correction at high $p_T$ is similar to the one already
found in $WZ$~\cite{Campanario:2012fk} and
$WW$~\cite{Campanario:2013wta} production.
We note that, above 200 GeV, the $R_\LS$ uncertainty is much smaller than the
uncertainty coming from variations of the factorization and renormalization
scales. Hence, the error related to assigning emission sequence by LoopSim
is negligible.
The overall theoretical uncertainty decreases by 20-30\% as we go from NLO to
\nNLO for these distributions.
We consider the \nNLO predictions shown in Fig.~\ref{fig:pTlmax-ptZlead} as
one of the highlights of our study. We emphasize that the inclusion of QCD
corrections of that size should be mandatory in all related diboson analyses at
high transverse momenta. Such large corrections are of huge importance and
should be accounted for both in computations of SM backgrounds and in
searches for anomalous couplings.
 
We notice that the kink around 900 \GeV in the $p_{T,Z}^\text{lead}$
distribution is due to the $\Delta R_{\ell, \ell}$ cut of Eq.~(\ref{eq:cutsincl}).
For a high-$p_T$ boson, the decay
products are highly collimated with the $Z$ transverse momentum and they share
approximately equal amounts of its $p_T$.  Hence, the mass of the dilepton
system can be approximated by $m^2_{\ell\ell} \simeq \frac14 p_{T,Z}^2 \Delta R^2_{\ell,\ell}$.
The imposed phase-space cut $\Delta R_{\ell,\ell} > 0.2$ then leads to the condition
$p_{T,Z} \lesssim 10 \, m_{\ell\ell}$. 
Since the $m_{\ell,\ell}$ distribution is strongly peaked at the $Z$ boson mass,
we obtain that the typical separation between the leptons drops below 0.2 at
$p_{T,Z} \simeq 900 \GeV$. This leads to the kink observed in the LO distribution of 
Fig.~\ref{fig:pTlmax-ptZlead}. Also, because the two $Z$ bosons are
back-to-back in LO configurations and hence have the same $p_T$, the
same effect happens simultaneously for both $Z$ bosons.
With additional parton radiation, this effect is smoothed out. The
sub-leading $Z$ boson will in general have a smaller $p_T$, as some
transverse momentum is carried by the parton.
The kink at about 800~GeV in the $p_{T,\ell}^\text{max}$ distribution,
shown in the right panel of Fig.~\ref{fig:pTlmax-ptZlead}, is also due
to the $\Delta R_{\ell,\ell}$ cut. We have checked
explicitly that changing the $\Delta R_{\ell,\ell}$ cut to 0.4 moves the
kink to 450 and 400~GeV for the $p_{T,Z}^\text{lead}$ and
$p_{T,\ell}^\text{max}$ distribution, respectively, consistent with the
above discussion.

\begin{figure}[t]
  \centering
  \includegraphics[width=0.45\columnwidth]{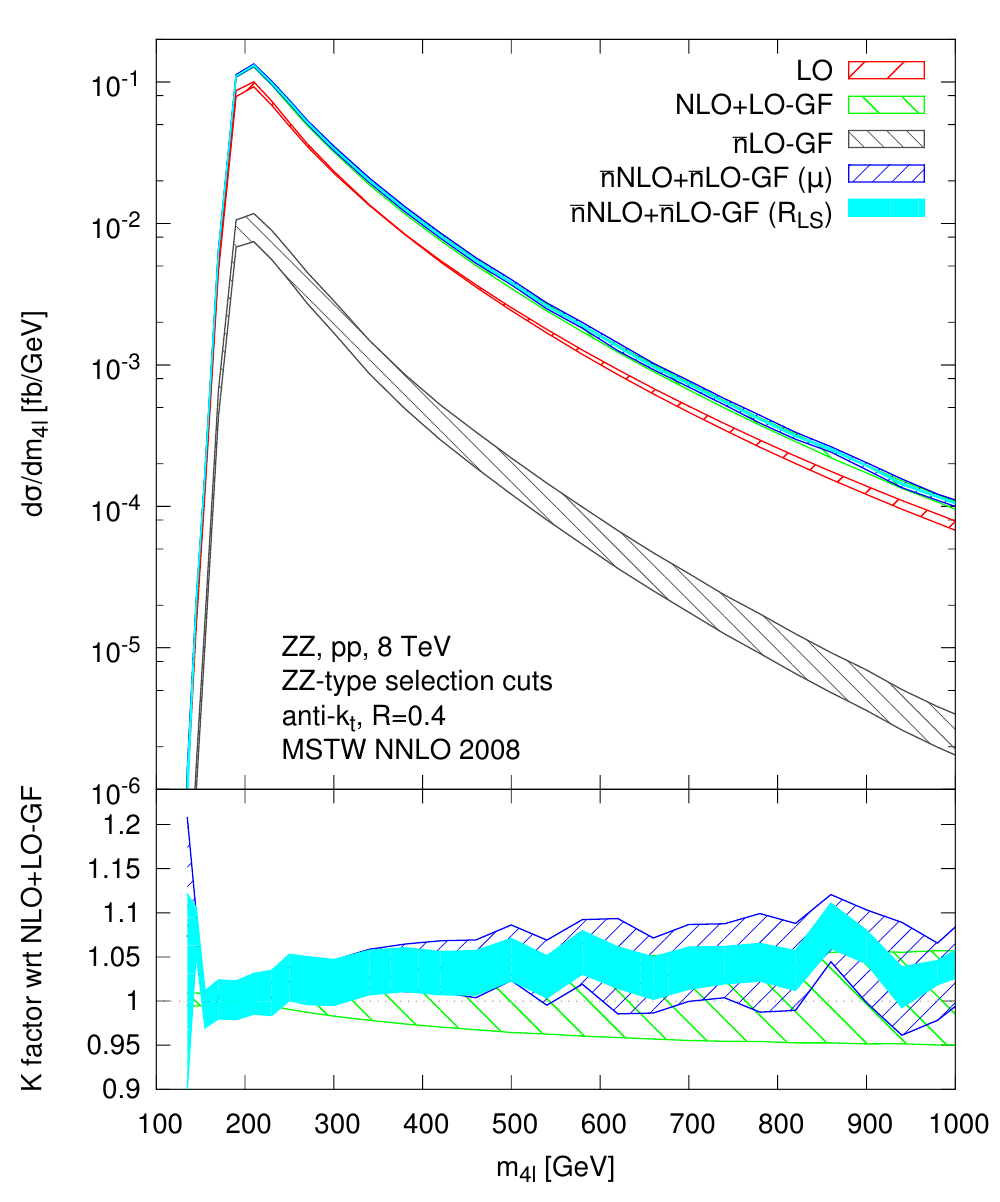}
  \hfill
  \includegraphics[width=0.45\columnwidth]{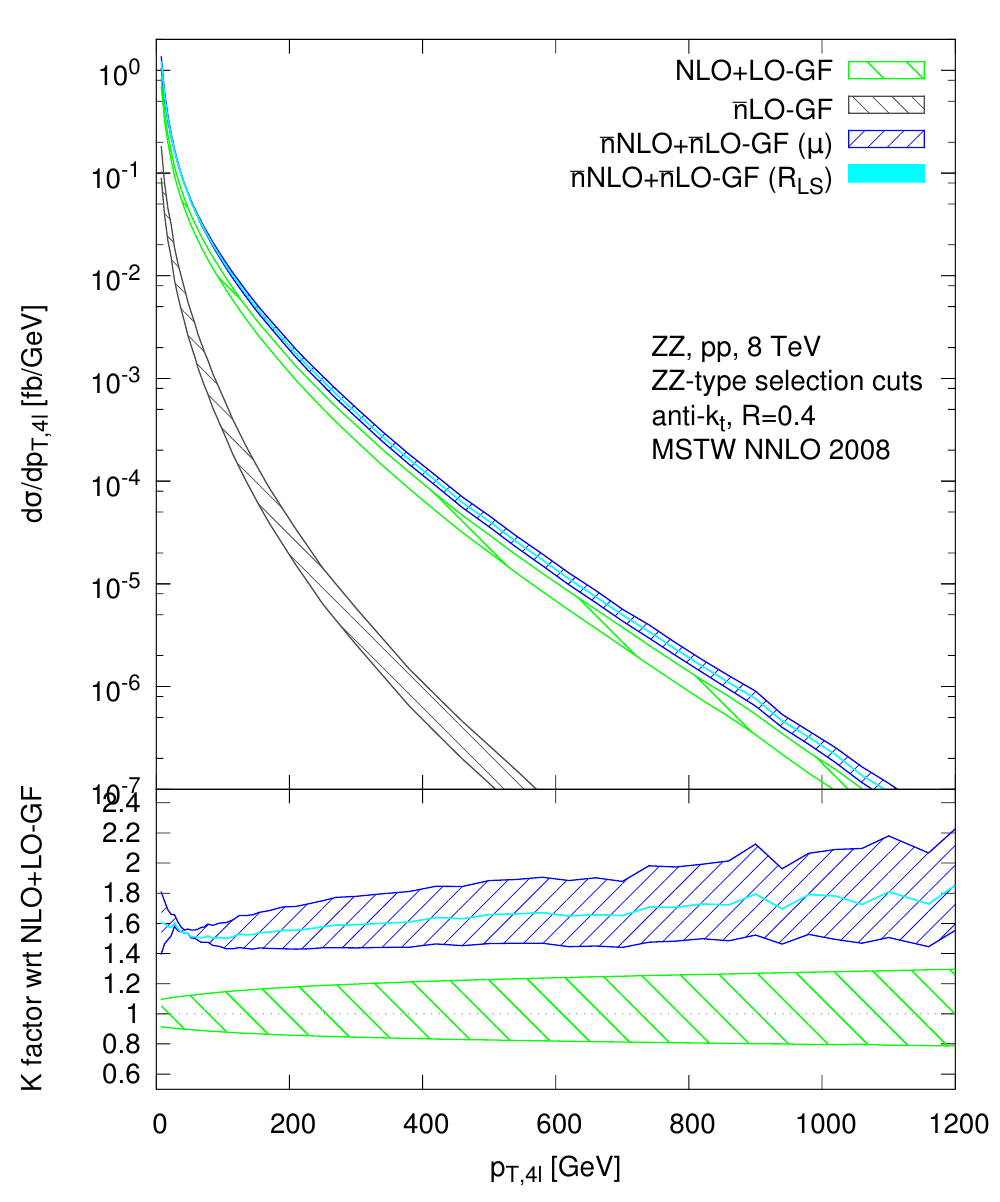}
  \caption{  
  Differential cross sections and K factors for invariant mass~(left) and the
  transverse momentum~(right) of the four-lepton system for the \zz-type
  selection of Eq.~(\ref{eq:mZselections}) at the LHC with $\sqrt{s}=8\,
  \text{TeV}$. 
  All details as in Fig.~\ref{fig:HT-standard}.
  } 
  \label{fig:ZZsys}
\end{figure}

Another interesting distribution is the mass of the \zz system or, equivalently,
the mass of the system of four leptons, $m_{4\ell}$, which is shown in
Fig.~\ref{fig:ZZsys}. This observable is particularly important from the point
of view of anomalous triple gauge coupling (aTGC) searches, as the effects of
anomalous couplings 
are enhanced in events where large momentum is transferred through the
triple-boson vertex~\cite{Aad:2012awa, Khachatryan:2014dia}. 
Such events result in a large invariant mass of the di-boson system.
It is interesting to find that the $m_{4\ell}$ distribution receives only modest
\nNLO corrections from QCD, which stay always below 5\%. This is to be compared
with typical sizes of electro-weak corrections, which are not accounted
for, and can be of the order of 10\% in the tail of the distributions,
and with the PDF uncertainties, estimated at 5-10\%.
Hence, compared to the above sources of uncertainties, the \nNLO QCD corrections
are small and we conclude that the $m_{4\ell}$ distribution becomes stable at
this order and can be safely used for setting aTGC limits~\cite{Aad:2012awa,
Khachatryan:2014dia}.

Fig.~\ref{fig:ZZsys} (right) shows the distribution of the transverse momentum of
the four-lepton system. At LO, this distribution is just a $\delta$-function at
$p_{T,4\ell} = 0$, since there is no jet the four-lepton system could recoil
against. We do not show this first bin in the figure. The first non-trivial order
for this observable is then NLO. As we go to \nNLO, the $p_{T,4\ell}$
distribution receives significant corrections of the order of 60-80\% for
the range presented in the plot, consistent with the NLO predictions shown
for $ZZ$+jet production in Ref.~\cite{Binoth:2009wk}. 

\begin{figure}[t]
  \centering
  \includegraphics[width=0.45\columnwidth]{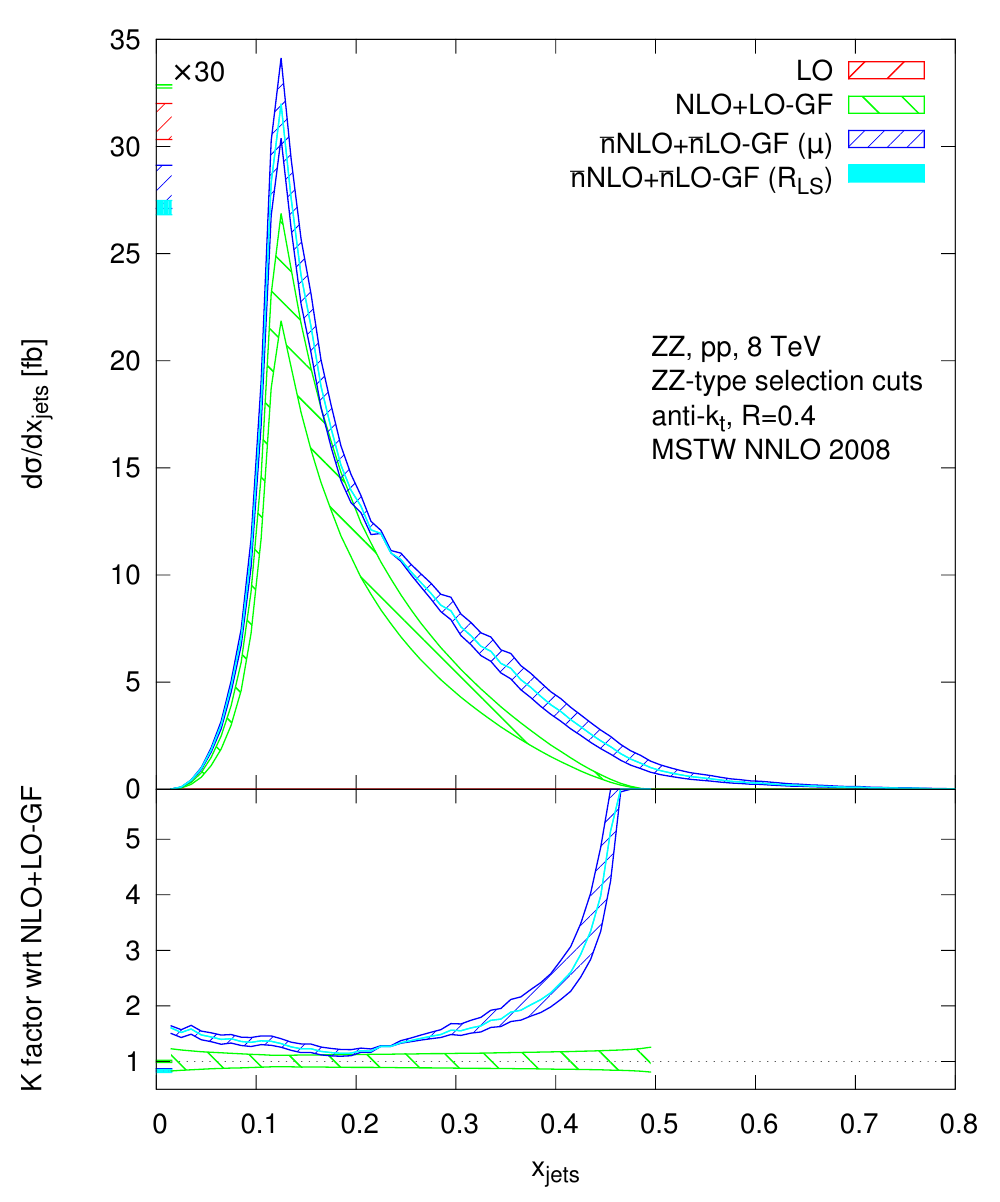}
  \caption{  
  Differential cross sections and K factors at various perturbative orders for 
  the $x_\jet$ variable defined in Eq.~(\ref{eq:xjet-def}).
  The results correspond to the LHC with $\sqrt{s}=8\, \text{TeV}$ and
  \zz-type selection of Eq.~(\ref{eq:mZselections}).
  All details as in Fig.~\ref{fig:HT-standard}.
  }
  \label{fig:xs}
\end{figure}

Finally, in Fig.~\ref{fig:xs}, we show the distribution of the $x_\jet$ variable
defined as
\begin{equation}
  x_\jet = \frac{\sum_{\substack{k \in \{{\text{jets}}}\}}E_{T,k}}
                {\sum_{\substack{k \in \{{\text{jets}, Z\text{s}}}\}}E_{T,k}}\,,
  \label{eq:xjet-def}
\end{equation}
which has been introduced in Ref.~\cite{Campanario:2014lza} in the context of a
dynamical jet veto. Large values of this variable correspond to configurations where
most of the energy of the final state is carried by jets. For the case of
$ZZ$ production, $x_\jet = 0$ at LO and the distribution starts to be non-trivial
at NLO. Here we now also show the first bin, which almost exclusively
consists of the $x_\jet = 0$ contributions, and which we have scaled
down by a factor 30 to fit the plot range. At LO, including LO-GF, the
result is simply the inclusive cross section of the process. At higher
orders, this bin corresponds effectively to the contribution after
placing a jet veto on the process. 
 
As we see in Fig.~\ref{fig:xs}, at NLO, where one jet is allowed, the
$x_\jet$ distribution is peaked around 0.1-0.15, hence, for most events, the jet
carries 10-15\% fraction of the final state energy. 
We also see, however, that there is a non-negligible tail reaching out to $x_\jet = 0.5$.
When we move to \nNLO, the yield of $x_\jet$ events increases over the entire range
of the distribution
at the expense of the energy carried by the $Z$ bosons. In particular, the tail
receives corrections with K factors of the order of 5 around
$x_\jet=0.5$. 
The \nNLO result suggests that the $x_\jet$ cut, used in dynamical jet veto
analyses~\cite{Campanario:2014lza}, should be set around $x_\jet = 0.2$,
somewhat lower than what could be inferred from the NLO distribution.
The $R_\LS$ uncertainty is negligible for this observable and the
renormalization and factorization scale uncertainty is comparable at NLO and
\nNLO.

\subsubsection{Higgs analysis}
\label{sec:hig}

\begin{figure}[t]
  \centering
  \includegraphics[width=0.45\columnwidth]{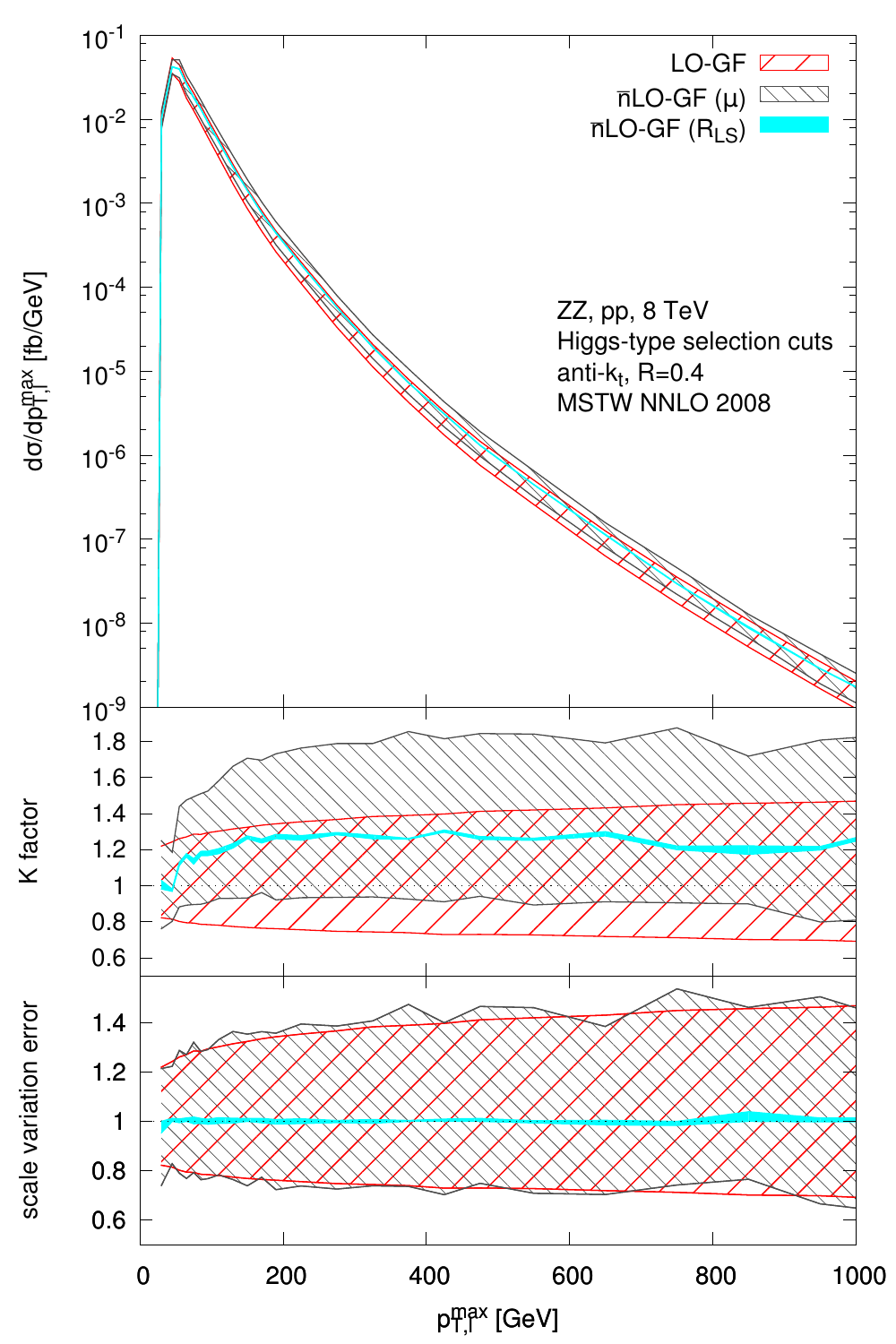}
  \hfill
  \includegraphics[width=0.45\columnwidth]{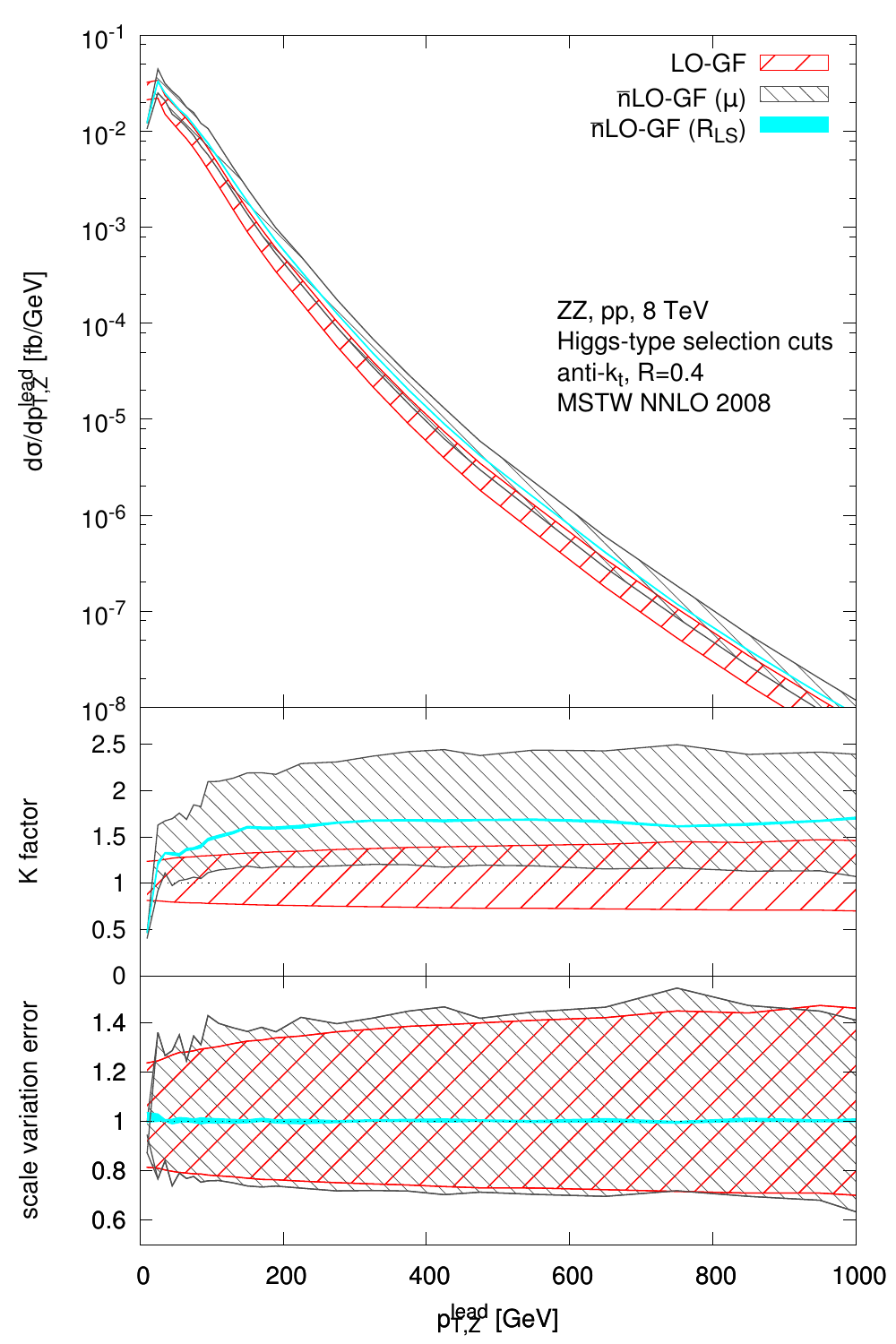}
  \caption{  
  Differential cross sections and K factors for transverse momentum of the
  leading lepton~(left) and the leading $Z$ boson~(right) for the  Higgs-type
  selection defined in Eqs.~(\ref{eq:Higgs-sel1}) and (\ref{eq:Higgs-sel2}).
  The results correspond to the LHC at $\sqrt{s}=8\, \text{TeV}$.
  All other details as in Fig.~\ref{fig:HT-standard}.
  }
  \label{fig:box-nLO}
\end{figure}

\begin{figure}[t]
  \centering
  \includegraphics[width=0.45\columnwidth]{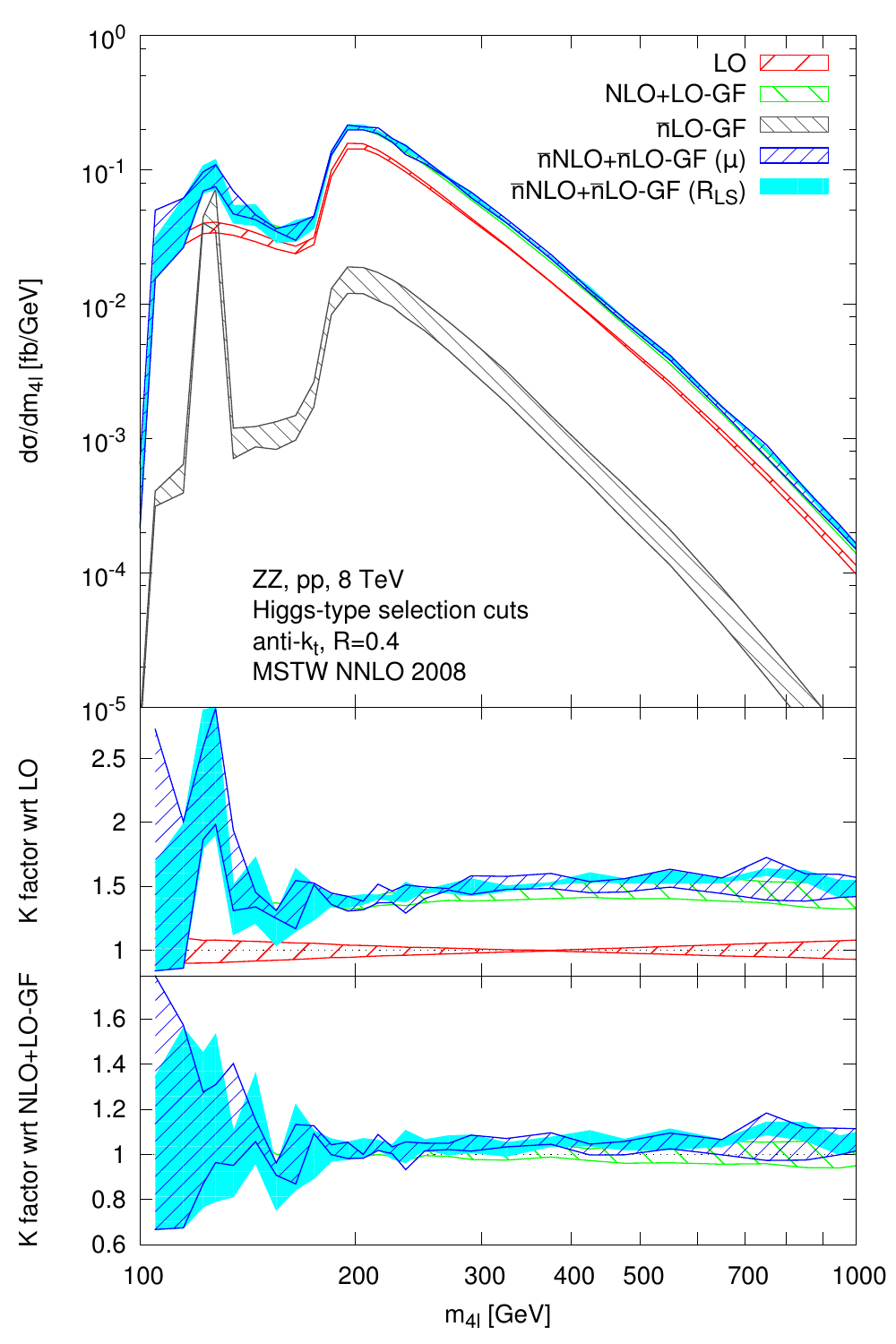}
  \caption{  
  Differential cross sections and K factors for the invariant mass of the
  four-lepton system for the  Higgs-type selection defined in
  Eqs.~(\ref{eq:Higgs-sel1}) and~(\ref{eq:Higgs-sel2}).  The results correspond
  to the LHC at $\sqrt{s}=8\, \text{TeV}$.
  All other details as in Fig.~\ref{fig:HT-standard}.
  }
  \label{fig:Higgs-sel-mZZ}
\end{figure}

We turn now to the discussion of the four-lepton production in the context of
Higgs analyses. As explained in Section~\ref{sec:theo} and shown in
Fig.~\ref{fig:feynmanZZ}, the same $ZZ \to 4\ell$ signature can be produced with
and without the intermediate Higgs boson. The latter constitutes an
irreducible background from the point of view of studies of Higgs production in 
gluon-gluon fusion, therefore, its precise determination is of utmost
importance. The analyses optimized for Higgs studies use slightly different set
of cuts with respect to those employed for aTGC searches.

Following the CMS analyses of Ref.~\cite{CMS:xwa}, where Higgs
properties are studied, we require
\begin{align}
  p_{t,e} & >  7 \GeV\,,    & |\eta_e|  <2.5\,, \nonumber \\
  \label{eq:Higgs-sel1}
  p_{t,\mu} & >  5 \GeV\,,  & |\eta_\mu|  <2.4\,, \\
  p_{t,\ell_\text{hardest}} &> 20 \GeV\,,  & m_{4\ell}>100 \GeV\,, \nonumber \\
  p_{t,\ell_\text{second-hardest}}&> 10 \GeV\,,  & \nonumber
\end{align}
and
\begin{eqnarray}
  40 < m_{\ell \ell} < 120 \GeV   & & \text{for the $\ell \ell$
  pair with mass closer to $m_Z$}, 
  \nonumber \\
  12 < m_{\ell \ell} < 120 \GeV   & & \text{for the other $\ell \ell$ pair}, 
  \label{eq:Higgs-sel2}
  \\
  m_{\ell\ell} > 4 \GeV & & \text{for any oppositely-charged pair of leptons}.
  \nonumber
\end{eqnarray}

We start by showing in Fig.~\ref{fig:box-nLO} the distributions of transverse
momenta of the leading $Z$ and the leading lepton, focusing only on the GF part.
The LO result corresponds exactly to the diagrams of
Fig.~\ref{fig:feynmanZZ}~(c) and~(d). The \nLO correction is computed with
LoopSim using the $gg \to 4\ell + j$ result of
Ref.~\cite{Campanario:2012bh} provided by VBFNLO.
We see that the correction to $p_{T,Z}^\text{lead}$ and $p_{T,\ell}^\text{lead}$
is large practically over the entire range shown in Fig.~\ref{fig:box-nLO}.  The
K factor reaches up to 50\% for central values and the scale variation does not
decrease as we go from LO to \nLO (bottom panel). It is important to notice that
the $R_\LS$ uncertainty is negligible for those distributions, hence our
prediction is very reliable within the uncertainty coming from the factorization
and renormalization scale.
 
Let us now turn to the distribution of the invariant mass of the $ZZ$ pair.  As argued
in Ref.~\cite{Kauer:2012hd}, even though the Higgs width is extremely small in the
Standard Model, interference effects between continuum and Higgs-mediated
contributions (diagrams (c) and (d) in Fig.~\ref{fig:feynmanZZ}) lead to
enhanced four-lepton mass spectra in the region $m_{4\ell} > 2 m_Z$.
Hence, the off-shell effects cannot be neglected, and, as shown
recently~\cite{Caola:2013yja, Campbell:2013una}, by comparing the yield at the
Higgs mass peak with that off the peak, they can be used for setting
bounds on the Higgs decay width.

Our predictions for the four-lepton mass spectra are shown in
Fig.~\ref{fig:Higgs-sel-mZZ}. Similarly to the $ZZ$ selection discussed in the
previous section, also here, \nNLO QCD effects bring a very small correction to
this observable. Hence, the $m_{4\ell}$ distribution is stable at this order.
This is important from the point of view of the off-shell effect studies in GF
fusion, since a precise determination of the expected theoretical yield has 
impact on setting the Higgs width limits.

\section{Summary}
\label{sec:su}
We have studied the $ZZ$ production process at the LHC beyond NLO in QCD by
merging the $ZZ$ and $ZZ$+jet NLO samples with help of the LoopSim method.  
We have included the exact, loop-induced GF predictions, which are part of 
NNLO, as well as the GF $ZZ$+jet contribution, which is formally of N$^3$LO.
The NLO samples were obtained with the VBFNLO package. 
The leptonic decays of the $Z$ bosons, including all 
off-shell and spin correlation effects, have been fully taken into account.
 
We have compared the fully inclusive cross section predictions from our
framework with the exact results computed in Ref.~\cite{Cascioli:2014yka} and
found a very good agreement within 2$\% $, covered by the remaining
scale uncertainties, despite the fact that the LoopSim method is missing
some finite parts originating from the two-loop virtual contributions.
Following closely two experimental setups, one used in the SM $ZZ$
production and aTGC searches, and the other used in Higgs analyses, we obtained
results for a selection of differential distributions.
 
For the observables sensitive to QCD radiation, the corrections exceed the
errors bars of the NLO predictions combined with the LO-GF results and range from 20$\%$, in the case of the
transverse momentum of the \z boson or the leading lepton, up
to 100$\%$ for the effective mass variable $H_{T}$.
A study of the $x_\text{jet}$ observable, defined as ratio of the transverse
energy sum of the jets over the sum of transverse energies of jets and
$Z$ bosons, shows that the \nNLO corrections start to increase quickly
when this variable exceeds $0.2$. Therefore, when this variable is used
to impose a dynamical jet veto, a cut should be placed at around this
value.
For Higgs-type selection, we investigated also the \zz production via GF at
\nLO. The size of the corrections are around 50$\%$ for the transverse 
momenta of the leading $Z$ and 20$\%$ for those of the leading lepton.
 
For observables which favor the LO kinematics, like $m_{4l}$, the approximated
NNLO QCD corrections are small, of the order of 5$\%$, and comparable
with the size of the remaining scale or PDF uncertainties.
 
Modifications to the VBFNLO program used in this article are available on
request and will be part of a future release.  The LoopSim library, together
with the Les Houches Event interface, is publicly available at
\url{https://loopsim.hepforge.org}.

\section*{Acknowledgments}

This work was performed on the computational resource bwUniCluster funded by
the Ministry of Science, Research and Arts and the Universities of the State
of Baden-W\"urttemberg, Germany, within the framework program bwHPC.
FC has been partially funded by a Marie Curie fellowship
(PIEF-GA-2011-298960), the Spanish
Government and ERDF funds from the EU Commission
[Grants No. FPA2011-23596, FPA2011-23778, FPA2014-53631-C2-1-P
No. CSD2007-00042 (Consolider Project CPAN)]
S.S. is grateful for hospitality to the Institut f\"ur Theoretische Physik at
the Karlsruhe Institute of Technology, where part of this work has been
done. 


\end{document}